\documentclass[lettersize,journal]{IEEEtran}
\usepackage{amsmath,amsfonts,amssymb}
\usepackage{multicol}
\usepackage{array} 
\usepackage{tabularray}
\usepackage{nicematrix}
\usepackage{booktabs}
\usepackage{colortbl}
\usepackage{ragged2e}
\usepackage{longtable}

\usepackage{multirow}
\usepackage{hhline}
\usepackage{makecell}
\usepackage{rotating}

\usepackage{algorithm}
\usepackage{algpseudocode}
\usepackage[caption=false,font=normalsize,labelfont=sf,textfont=sf]{subfig}
\usepackage{textcomp}
\usepackage{stfloats}
\usepackage{url}
\usepackage{verbatim}
\usepackage{graphicx}
\usepackage{cite}
\usepackage{pdfpages}
\newtheorem{theorem}{Theorem}[section]

\newtheorem{lemma}[theorem]{Lemma}
\newtheorem{definition}{Definition}
\newtheorem{proposition}[theorem]{Proposition}

\newtheorem{example}{Example}
\newtheorem{remark}{Remark}

\newenvironment{prof}{\textit{Proof\,:}} { $\square$}

\usepackage[utf8]{inputenc}
\begin{document}
	
    \title{Full-Diversity Construction-D Lattices: Design and Decoding Perspective on Block-Fading Channels}
	
	\author{{\uppercase{Maryam Sadeghi},
			\IEEEmembership{Member, IEEE},
			, \uppercase{Hassan Khodaiemehr}, and \uppercase{Chen Feng}, \IEEEmembership{Member, IEEE}}}
	
	\markboth{IEEE Transactions on Communications,~Vol.~X, No.~X, Nov~2024}%
	{Sadeghi \MakeLowercase{\textit{et al.}}: Full-Diversity Construction-D Lattices: Design and Decoding Perspective on Block-Fading Channels}

\author{\IEEEauthorblockN{Maryam Sadeghi~\IEEEmembership{Member, IEEE},  Hassan Khodaiemehr and Chen Feng~\IEEEmembership{Member, IEEE}}
\thanks{
M. Sadeghi, H. Khodaiemehr, and C. Feng are with the School of Engineering, Faculty of Applied Science, University of British Columbia (UBC), Okanagan Campus, Kelowna, BC, Canada,
(e-mails: maryam.sadeghi@ubc.ca, hassan.khodaiemehr@ubc.ca and chen.feng@ubc.ca).
\\
Some of the findings from this paper have been submitted to ISIT 2025 \cite{ISIT2025}.
}}
	
	
	\maketitle
	
	\begin{abstract}
   This paper introduces a novel framework for constructing algebraic lattices based on Construction-D, leveraging nested linear codes and prime ideals from algebraic number fields. We focus on the application of these lattices in block-fading (BF) channels, which are characterized by piecewise-constant fading across blocks of transmitted symbols. This approach results in a semi-systematic generator matrix, providing a structured foundation for high-dimensional lattice design for BF channels. The proposed Construction-D lattices exhibit the full diversity property, making them highly effective for error performance improvement.  To address this, we develop an efficient decoding algorithm designed specifically for full-diversity Construction-D lattices. 
   Simulations indicate that the proposed lattices notably enhance error performance compared to full-diversity Construction-A lattices in diversity-2 cases. Interestingly, unlike AWGN channels, the expected performance enhancement of Construction-D over Construction-A, resulting from an increased number of nested code levels, was observed only in the two-level and diversity-2 cases. This phenomenon is likely attributed to the intensified effects of error propagation that occur during successive cancellation at higher levels, as well as the higher diversity orders.
   These findings highlight the promise of Construction-D lattices as an effective  coding strategy for enhancing communication reliability in BF channels.
	\end{abstract}
	
	\begin{IEEEkeywords}
		Block fading channels, Construction-D lattices, algebraic number fields.
	\end{IEEEkeywords}
\section{Introduction}
\IEEEPARstart{I}{n} numerous applications, including sphere packing, quantization, and modulation, lattices, which are discrete subgroups of $\mathbb{R}^n$, provide a powerful framework.  Their use in coding for additive white Gaussian noise (AWGN) channels has been extensively explored, with infinite lattices often serving as unrestricted codes optimized for AWGN performance \cite{Poltyrev}.  The capacity-achieving potential of lattice codes on AWGN channels is proven, with the existence of codes exceeding any rate below capacity, while maintaining an average error probability below a specified tolerance \cite{urbanke,erez2004achieving}. Shaping region selection allows transforming capacity-achieving lattices into corresponding lattice codes \cite{erez2004achieving}.



A variety of methods exist for constructing lattices, including Constructions A, B, C, D, and D$'$, each leveraging distinct code types to achieve specific properties relevant to communication systems \cite{conway1998nja}.  Key distinctions arise from the underlying codes: Construction B, for example, utilizes even-weight binary codes, defining lattice points as vectors $\mathbf{x} = (x_1, x_2, \dots, x_n)$ where $\sum_{i=1}^n x_i$ is a multiple of four.  Constructions A and C offer greater flexibility in code selection. While Construction C achieves optimal sphere packing in $\mathbb{R}^n$, it does not always result in a lattice structure.


Construction-D stands out for its hierarchical approach to lattice generation using nested linear codes \cite{barnes1983new}. This multi-level structure provides precise control over lattice parameters, making it highly suitable for coding theory, cryptography, and communications.  A significant result demonstrates that Construction-D lattices, along with a broad class of multilevel coset codes, nearly achieve the sphere packing bound \cite{forney2000sphere}. This finding, based on information-theoretic channel coding, highlights the volume-to-noise ratio (VNR) as a key performance metric.  A major goal in lattice design is finding structured lattices that balance practical complexity with near-sphere-packing performance, enabling transmission with arbitrarily low error probabilities as VNR approaches unity, thus facilitating capacity-achieving lattices and codes.

The inherent complexity of maximum-likelihood (ML) decoding for high-dimensional lattices necessitates the development of low-complexity alternatives.  Soft-decision decoding applied to underlying codes enables linear-complexity decoding of Construction-A, D, and D$'$ lattices \cite{1057135,LDLC,LDA,LDA2,LDA3,Leech,polar,QC_LDPC_lattice}.  Construction-D lattices, constructed using various nested linear codes, including turbo and BCH codes, have been extensively studied.  Matsumine \emph{et al.} \cite{matsumine2018construction} presented a modified multistage decoding algorithm for BCH-code-based Construction-D lattices, demonstrating superior performance of dimension-128 lattices (using ordered statistics decoding) compared to Construction-D$'$ and turbo-code-based Construction-D counterparts.  Furthermore, the utilization of polar codes within Construction-D yields lattices outperforming Barnes-Wall lattices in terms of symbol error rate over AWGN channels and achieving AWGN capacity \cite{Liu}.

While practical lattice coding over AWGN channels is well understood, its generalization to other channel types, particularly block-fading (BF) channels, remains less explored \cite{blockfading}.  BF channels, relevant to systems employing frequency-hopping or orthogonal frequency division multiplexing (OFDM), exhibit non-ergodic behavior resulting in zero Shannon capacity \cite{rootLDPC}.  Algebraic lattices, constructed from the ring of integers of a number field, offer promising modulation schemes for fading channels \cite{8187356} and represent a natural candidate for BF channel coding due to their full-diversity properties. Although Construction-A lattices have been generalized to various settings including cyclotomic fields \cite{ebeling} and number fields \cite{ConstA},  their error performance over BF channels has the potential for further improvements~\cite{Hassan}.

When extended to algebraic settings, such as over the ring of integers \(\mathcal{O}_K\) of a number field \(K\), Construction-D lattices display additional algebraic structures that can enhance performance in diverse applications. Previous studies on Construction-D algebraic lattices have extended the original construction, as well as Forney’s code formula, from binary codes to linear codes over finite rings \(\mathbb{Z}_q\) with \(q > 2\) \cite{strey2017lattices}. A key contribution includes the introduction of a zero-one addition operation in \(\mathbb{Z}_q\), which coincides with the Schur product for binary codes. It was demonstrated that these extended constructions form lattices if and only if the nested codes are closed under this addition. These generalizations are particularly important due to the growing use of \(q\)-ary lattices in cryptographic applications. A recent study investigates key properties of lattices obtained through Constructions D and D$'$ using \(q\)-ary linear codes, exploring their connections with Construction-A, as well as deriving their generator matrices, volume expressions, and bounds on minimum distances \cite{do2023lattice}.

Algebraic lattices, known for achieving full diversity, which is a crucial property in fading channels, have seen recent advancements \cite{Hassan}.  These advances focus on constructing full-diversity algebraic lattices for  BF channels using Construction-A over totally real number fields.  The proposed lattices utilize two linearly scalable decoding algorithms.  The first algorithm, designed for algebraic low-density parity-check (LDPC) lattices, combines iterative and non-iterative phases leveraging a specialized parity-check matrix and Tanner graph for efficient iterative decoding.  This iterative approach, when combined with LDPC codes achieving the outage probability limit over single-block fading channels, allows these lattices to achieve a full diversity order of $n$ over an $n$-block BF channel.  A simplified, non-iterative decoding variant is also presented, providing a more practical approach to full-diversity decoding of generic Construction-A lattices.


This paper introduces novel algebraic Construction-D lattices defined over the ring of integers $\mathcal{O}_K$, analyzing their properties and performance in lattice-based coding schemes for BF channels. Our results demonstrate a strong correlation between lattice performance and both the algebraic properties of $\mathcal{O}_K$ and the number of nested codes employed in the Construction-D framework. Our approach leverages the interplay between the algebraic properties of $\mathcal{O}_K$ and the multi-level characteristics of Construction-D to enhance error correction capabilities. In comparison to prior work using Construction-A over totally real number fields \cite{Hassan}, we present a semi-systematic generator matrix for Construction-D lattices and a successive cancellation decoding algorithm, demonstrating improved error performance over BF channels while maintaining full diversity under the conditions outlined in \cite{Hassan}.

The organization of this paper is as follows: Section~\ref{pre} presents the preliminaries. In Section~\ref{ConD}, we introduce Construction-D lattices based on algebraic number fields. Section~\ref{alg} discusses the application of algebraic lattices in block fading channels. In Section~\ref{decoding}, we propose our full-diversity decoding method, characterized by linear complexity in the lattice dimension, along with its diversity analysis. Section~\ref{result} presents the simulation results and performance comparisons with Construction-A lattices in the context of block fading channels. The future research directions are also outlined in this section. Finally, Section~\ref{conclusion} offers our concluding remarks.

\textbf{Notation}: Matrices are represented by bold uppercase letters, while vectors are represented by bold lowercase letters. The $i^{th}$ coThe future research directions are also outlined in this section. ponent of the vector $\mathbf{a}$ is indicated by $a_i$ or $\mathbf{a}(i)$, and the matrix $\mathbf{A}$ has its entry at position $(i,j)$ represented as $A_{i,j}$ or $\mathbf{A}(i,j)$. The notation $[\cdot]^t$ refers to the transpose of vectors and matrices. For a vector $\mathbf{x}$ of length $n$, where $1 \leq i < j \leq n$, the notation $\mathbf{x}(i:j)$ is used throughout this paper to denote the subvector of $\mathbf{x}$ containing the elements from $i$ to $j$.
\section{Preliminaries}\label{pre}
In this section, we establish the fundamental concepts essential for the forthcoming analysis of algebraic lattice constructions in the context of BF channels. The theory of lattices, which is deeply entrenched in algebraic number theory, offers a robust framework for tackling issues in coding and communication, especially in situations where reliability and efficiency are of utmost importance.
\subsection{Algebraic Number Fields}
Let $K$ be a subfield of $\mathbb{C}$ such that $[K:\mathbb{Q}]$ is finite, then $K$ is called a number field. Let $\theta\in K$ be a root of a non-zero polynomial $f(x)$ in $\mathbb{Q}[x]$, then $\theta$ is algebraic over $\mathbb{Q}$. Additionally, if the coefficients of $f(x)$ are in $\mathbb{Z}$, then $\theta$ is called an algebraic integer, and $\mathbb{Q}(\theta)$ is the smallest subfield of $K$ containing $\theta.$ Let $\mathcal{B}$ is the set of all algebraic integers which is a subring of $\mathbb{C},$ then $\mathcal{O}_K=K\cap\mathcal{B}$ is called the ring of integers of $K.$
\begin{definition}
	Let \( K = \mathbb{Q}(\theta) \) be a number field of degree \( n \) over \( \mathbb{Q} \). Then, \( K \) has precisely \( n \) embeddings \( \sigma_1, \ldots, \sigma_n \) into \( \mathbb{C} \), each defined by \( \sigma_i(\theta) = \theta_i \) for \( i = 1, \ldots, n \), where \( \theta_i \)'s are the distinct roots of the minimal polynomial of \( \theta \) over \( \mathbb{Q} \) in \( \mathbb{C} \).
\end{definition}
For a number field $K$, the ordered pair $(r_1,r_2)$ where $r_1$ is the number of real embeddings of $K$ and $r_2$ is the number of complex conjugate pairs of embeddings, is called the signature of $K.$ The degree of $K$ is defined as $n=r_1+2r_2$. If $r_2=0$, we have a totally real number field and if $r_1=0$, we have a totally complex number field. 
We arrange the embeddings \( \sigma_i \) so that, for all \( x \in K \), \( \sigma_i(x) \in \mathbb{R} \) for \( 1 \leq i \leq r_1 \), and \( \sigma_{j+r_2}(x) \) is the complex conjugate of \( \sigma_j(x) \) for \( r_1 + 1 \leq j \leq r_1 + r_2 \). The canonical embedding \( \sigma : K \rightarrow \mathbb{R}^{r_1} \times \mathbb{C}^{r_2} \) is the homomorphism defined by
\begin{align}\label{canon}
	\sigma(x) = \left( \sigma_1(x), \ldots, \sigma_{r_1}(x), \sigma_{r_1+1}(x), \ldots, \sigma_{r_1+r_2}(x) \right).
\end{align}
When we identify \( \mathbb{R}^{r_1} \times \mathbb{C}^{r_2} \) with \( \mathbb{R}^n \), the canonical embedding can be expressed as
\[
\sigma : K \rightarrow \mathbb{R}^n,
\]
\begin{equation}
	\begin{split}
		\sigma(x) = \big(\sigma_1(x), \ldots, \sigma_{r_1}(x),
		\Re(\sigma_{r_1+1}(x)), \Im(\sigma_{r_1+1}(x)), \\
		\ldots, \Re(\sigma_{r_1+r_2}(x)), \Im(\sigma_{r_1+r_2}(x)) \big),
	\end{split}
\end{equation}
where \( \Re \) denotes the real part and \( \Im \) denotes the imaginary part.
 \begin{theorem}[\emph{\cite{485720}}]
 For a number field \(K\) with the ring of integers \(\mathcal{O}_K\) and signature $(r_1,r_2)$, the algebraic lattice of the form \(\sigma(\mathcal{O}_K)\) exhibits a diversity order \(L = r_1 + r_2\).
 \end{theorem}
 Given that \( r_1 + 2r_2 = n = [K: \mathbb{Q}] \) and for totally real number fields \( r_2 = 0 \), algebraic lattices derived from such fields achieve a diversity order of \( n \), meaning they are full-diversity lattices.

\begin{definition}
Let \( K \) be a number field of degree \( n \) and \( x \in K \). The elements \( \sigma_1(x), \ldots, \sigma_n(x) \) are the conjugates of \( x \). The norm and trace of \( x \) are given by
\[
\text{N}_{K/\mathbb{Q}}(x) = \prod_{i=1}^{n} \sigma_i(x), \quad \text{Tr}_{K/\mathbb{Q}}(x) = \sum_{i=1}^{n} \sigma_i(x),
\]
respectively. 
\end{definition}
For any \( x \in K \), both \( \text{N}_{K/\mathbb{Q}}(x) \) and \( \text{Tr}_{K/\mathbb{Q}}(x) \) are elements of \( \mathbb{Q} \). If \( x \in \mathcal{O}_K \), then both \( \text{N}_{K/\mathbb{Q}}(x) \) and \( \text{Tr}_{K/\mathbb{Q}}(x) \) are integers.
\begin{definition}
	Let $\{\alpha_1,\alpha_2,\ldots,\alpha_n\}$ form an integral basis for $K$ over $\mathbb{Q}$. Then, for $i,j=1,2,\ldots,n$, the discriminant of $K$ is defined as 
	\begin{align}
		\Delta_K=\det[\sigma_i(\alpha_j)]^2.
	\end{align}
	
\end{definition}
\begin{definition}[Monogenic Number Fields]
A number field \( K \) is called \textit{monogenic} if its ring of integers \( \mathcal{O}_K \) can be generated by a single element, forming a power basis. This means there exists an element \( \alpha \in \mathcal{O}_K \) such that any element \( \beta \in \mathcal{O}_K \) can be uniquely expressed as
\[
\beta = a_0 + a_1 \alpha + a_2 \alpha^2 + \cdots + a_{n-1} \alpha^{n-1},
\]
where \( a_i \in \mathbb{Z} \) for \( 0 \leq i < n \). In other words, we can write \( \mathcal{O}_K = \mathbb{Z}[\alpha] \) for some \( \alpha \in \mathcal{O}_K \), and the set \( \{1, \alpha, \alpha^2, \ldots, \alpha^{n-1}\} \) serves as a basis for \( \mathcal{O}_K \) over \( \mathbb{Z} \).

\end{definition}

\begin{definition}[Quadratic Number Fields]
	A quadratic number field \( K \) is a number field of degree 2 over \( \mathbb{Q} \), meaning \( [K : \mathbb{Q}] = 2 \). It can be expressed as \( K = \mathbb{Q}(\sqrt{m}) \), where \( m \) is a squarefree integer. When \( m < 0 \), \( K \) is called an imaginary quadratic field.
\end{definition}

The ring of integers  of \( K=\mathbb{Q}(\sqrt{m}) \) depends on the value of \( m \). If \( m \equiv 2 \) or \( 3 \pmod{4} \), then \( \mathcal{O}_K = \mathbb{Z}[\sqrt{m}] \), with a basis \( \{1, \sqrt{m}\} \) over \( \mathbb{Z} \) and a discriminant \(\Delta_K = 4m\). However, if \( m \equiv 1 \pmod{4} \), then the ring of integers is given by \( \mathcal{O}_K = \mathbb{Z}\left[ \frac{1 + \sqrt{m}}{2} \right] \) with discriminant \(\Delta_K = m\). In special cases, if \( m = -1 \), the field \( K \) yields the Gaussian integers, and if \( m = -3 \), it yields the Eisenstein integers.
\begin{theorem}[\emph{\cite{Alaca}}]
	Let \( a \) and \( b \) be integers such that the polynomial \( x^3 + ax + b \) is irreducible. Let \( \theta \in \mathbb{C} \) be a root of \( x^3 + ax + b \), and define the cubic field \( K = \mathbb{Q}(\theta) \) with \( \theta \in \mathcal{O}_K \). The discriminant \( \Delta_K\) of \( K \) over \( \mathbb{Q} \) is given by
	\[
	\Delta_K = -4a^3 - 27b^2.
	\]
	If the discriminant \( \Delta_K \) is square-free, or if \( \Delta_K = 4m \) for a square-free integer \( m \), where \( m \equiv 2 \) or \( 3 \pmod{4} \), then the set \( \{1, \theta, \theta^2\} \) forms an integral basis for the cubic field \( \mathbb{Q}(\theta) \).
\end{theorem}
\begin{theorem}
	Let \( a \) and \( b \) be integers such that the polynomial \( x^4 + ax + b \) is irreducible. Suppose \( \theta \in \mathbb{C} \) is a root of \( x^4 + ax + b \), and define the quartic field \( K = \mathbb{Q}(\theta) \) with \( \theta \in \mathcal{O}_K \). The discriminant of \( K \) over \( \mathbb{Q} \) is given by
	\[
	\Delta_K = -27a^4 + 256b^3.
	\]
	If  \( \Delta_K \) is square-free, then the set \( \{1, \theta, \theta^2, \theta^3\} \) forms an integral basis for the field \( \mathbb{Q}(\theta) \).
	
\end{theorem}
 
Let $\mathcal{P}$ be a prime ideal of $\mathcal{O}_{\mathbb{K}}$ and $p$ be a prime integer. We say $\mathcal{P}$ lies above $p$ if $\mathcal{P}|p\mathbb{Z}.$ Now, since $\mathcal{O}_K$ is a Dedekind domain, the ideal $p\mathcal{O}_K$ can be uniquely factorized as $p\mathcal{O}_K=\prod_{i=1}^m\mathcal{P}_i^{e_l}$, where $\mathcal{P}_i$'s are distinct prime ideals of $\mathcal{O}_K$ \cite{serglang,Stewart}. We call $e_l$ the ramification index of $\mathcal{P}_l$ over $p$ and $f_l=\left[\dfrac{\mathcal{O}_K}{\mathcal{P}_l}:\dfrac{\mathbb{Z}}{p\mathbb{Z}}\right]$ the inertia degree of $\mathcal{P}_l$ over $p$. 
The norm of an ideal \( \mathfrak{a} \) in a number field \( K \) is defined as 
$
N(\mathfrak{a}) = \left| \frac{\mathcal{O}_K}{\mathfrak{a}} \right|,
$
which represents the cardinality of the quotient group \( \mathcal{O}_K / \mathfrak{a} \).
Using this definition, it can be seen that every prime ideal $\mathcal{P}$ of $\mathcal{O}_K$ is maximal with norm $N(\mathcal{P})=p^f$, where $f\in\{1,2\}$ is the inertia degree, and we have $\dfrac{\mathcal{O}_K}{\mathcal{P}}\cong\mathbb{F}_{p^f}$.
\subsection{Lattices}
An $n$ dimensional lattice $\Lambda$ is a discrete subgroup of the Euclidean space $\mathbb{R}^n$ with vector addition operation which means that for any two lattice points $\mathbf{\lambda}_1$ and $\mathbf{\lambda}_2$ in $\Lambda$, $\mathbf{\lambda}_1+\mathbf{\lambda}_2$ and $\mathbf{\lambda}_1-\mathbf{\lambda}_2$ are in $\Lambda$. Every lattice \( \Lambda \) has a basis \( \mathbf{B} = \{ \mathbf{b}_1, \dots, \mathbf{b}_n \} \subseteq \mathbb{R}^m \), where \( n \leq m \) and the basis vectors \( \mathbf{b}_i \) are linearly independent. Each point \( \mathbf{x} \in \Lambda \) can be expressed as an integer linear combination of the vectors in \( \mathbf{B} \). The \( n \times m \) matrix \( \mathbf{G} \), with \( \mathbf{b}_1, \dots, \mathbf{b}_n \) as its rows, serves as a generator matrix for the lattice. The rank of the lattice is \( n \), and its dimension is \( m \). If \( n = m \), the lattice is referred to as a full-rank lattice. A lattice \( \Lambda \) can be defined by a generator matrix \( \mathbf{G} \) as follows:
\begin{equation}
	\Lambda=\{\mathbf{a}\mathbf{G}:\mathbf{a}\in\mathbb{Z}^n\}.
\end{equation}
The Voronoi cell of a lattice \( \Lambda \) in \( \mathbb{R}^m \) is defined as the set of all points in the space that are closer to a given lattice point \( \mathbf{v} \in \Lambda \) than to any other lattice point. Formally, the Voronoi cell \( \mathcal{V}(\mathbf{v}) \) associated with the lattice point \( \mathbf{v} \) is defined as:
\begin{multline}
	\mathcal{V}(\mathbf{v}) = \left\{ \mathbf{x} \in \mathbb{R}^m \mid \|\mathbf{x} - \mathbf{v}\| \leq \|\mathbf{x} - \mathbf{w}\| \right. \\
	\left. \text{for all } \mathbf{w} \in \Lambda \text{ with } \mathbf{w} \neq \mathbf{v} \right\}.
\end{multline}
\begin{definition}
	The discriminant of the lattice \( \Lambda \), denoted by \( \Delta(\Lambda) \), is defined as the determinant of the Gram matrix associated with any basis \( \mathbf{B} \) for $\Lambda$. Specifically, if \( \mathbf{G} \) is an \( n \times m \) generator matrix for \( \Lambda \), then the discriminant is  $\Delta(\Lambda) = \det(\mathbf{G}\mathbf{G}^t)$ and 
    \( \mathbf{G} \mathbf{G}^t\) is the Gram matrix of the lattice.
\end{definition}
The discriminant is related to the volume of the lattice by:
\begin{align}
\text{vol}(\Lambda) = \sqrt{\Delta(\Lambda)}.
\end{align}
The canonical embedding defined in Equation \eqref{canon} provides a geometric representation of a number field and establishes a connection between algebraic number fields and lattices.
Let \( \{ \omega_1, \dots, \omega_n \} \) be an integral basis of a number field \( K \). The \( n \) vectors \( \mathbf{v}_i = \sigma(\omega_i) \in \mathbb{R}^n \), for \( i = 1, \dots, n \), are linearly independent and thus define a full-rank algebraic lattice \( \Lambda = \Lambda(\mathcal{O}_K) = \sigma(\mathcal{O}_K) \).  If \( \Delta_K \) denotes the discriminant of the number field \( K \) and $r_2$ denotes the number of imaginary embeddings, then the volume of the fundamental parallelotope of \( \Lambda(\mathcal{O}_K) \) is given by \cite{8187356}:
\begin{align}
	\text{vol}(\Lambda(\mathcal{O}_K)) = 2^{-r_2} \sqrt{|\Delta_K|}.
\end{align}
\section{Construction-D  Lattices over Ring of Integers}\label{ConD}
Construction-D serves as a natural multi-level extension of Construction-A lattices. Construction-A lattices based on number fields are generated from the ring of integers \(\mathcal{O}_K\) of a number field \(K\). This construction begins by selecting an $[N,k]$-linear code $\mathcal{C}$ over $\mathbb{F}_p$, where $p$ is a prime number, and a non-zero prime ideal \(\mathfrak{p}\) in \(\mathcal{O}_K\), such that $\mathcal{O}_K/\mathfrak{p}\cong \mathbb{F}_p$. The aim is to create a lattice \(\Lambda\) that can be associated with $\mathcal{C}$.
The lattice is defined as follows:
\begin{eqnarray}\label{CostructionA}
\Lambda_{\mathcal{C}} = \left\{ \mathbf{x} \in \mathcal{O}_K^N : \mathbf{x} \equiv \mathbf{c} \pmod{\mathfrak{p}},  \mathbf{c} \in \mathcal{C}\subset \mathbb{F}_p^N \right\}.
\end{eqnarray}
\begin{theorem}{\cite[Proposition 1]{ConstA}}\label{theorem6}
The algebraic lattice \( \Lambda_{\mathcal{C}} \) in \eqref{CostructionA} is a sublattice of \( \mathcal{O}_K^N \) with a discriminant given by
\begin{equation}\label{disc}
  \Delta(\Lambda_{\mathcal{C}}) = \Delta_K^N (p^f)^{2(N-k)},
\end{equation}
where \( \Delta_K = (\det([\sigma_i(\omega_j)]_{i,j=1}^n))^2 \) represents the discriminant of the field \( K \). The lattice \( \Lambda_{\mathcal{C}} \) can be represented using the generator matrix
\begin{equation}\label{Gamma_C_gen}
  \mathbf{M}_{\Lambda} = \left[
                 \begin{array}{cc}
                   \mathbf{I}_k \otimes \mathbf{M} & \mathbf{A} \otimes \mathbf{M} \\
                   \mathbf{0}_{n(N-k) \times nk} & \mathbf{I}_{N-k} \otimes \mathbf{DM} \\
                 \end{array}
               \right],
\end{equation}
where \( \otimes \) denotes the tensor product of matrices. The matrix \( \left[
\begin{array}{cc}
 \mathbf{I}_k & \mathbf{A} \\
 \end{array}
  \right] \) serves as a generator matrix for the code \( \mathcal{C} \), while \( \mathbf{M} \) corresponds to the matrix of embeddings of a \( \mathbb{Z} \)-basis of \( \mathcal{O}_K \), and \( \mathbf{DM} \) is the matrix of embeddings of a \( \mathbb{Z} \)-basis of the prime ideal \( \mathfrak{p} \).
\end{theorem}
\subsection{Real Construction-D Lattices}	
Let $p$ be a prime in $\mathbb{Z}.$ Construction-D generates a lattice \( \Lambda \) from a sequence of nested linear codes 
\[
\mathcal{C}_1 \subset \mathcal{C}_2 \subset \cdots \subset \mathcal{C}_a \subseteq \left(\dfrac{\mathbb{Z}}{\langle p\rangle}\right)^N,
\]
where each \( \mathcal{C}_i \) is a linear code with parameters $[N,k_i]$ over $\dfrac{\mathbb{Z}}{\langle p\rangle}$ for \( i = 1, 2, \ldots, a \). Let \( \mathbf{M}_a \) be a generator matrix for the largest code \( \mathcal{C}_a \), structured such that it contains the generator matrices of all smaller codes \( \mathcal{C}_i \) for \( i=1,\ldots, a \). This can be represented as:
\[
\mathbf{M}_a = 
\begin{bmatrix}
	\mathbf{M}_1^t &
	\mathbf{M}_2'^t &
	\cdots &
	\mathbf{M}_{a-1}'^t &
	\mathbf{M}_a'^t
\end{bmatrix}^t,
\]
where \( \mathbf{M}_1 \) is a generator matrix for the code \( \mathcal{C}_1 \), and each \( \mathbf{M}_i' \), for \( i = 2, \ldots, a \), represents the additional rows in \( \mathbf{M}_a \) that are necessary to extend the generator matrices of the smaller code \( \mathcal{C}_i \) to its larger subsequent code \( \mathcal{C}_{i+1} \). Thus, the rows of each \( \mathbf{M}_i \) are included in the rows of \( \mathbf{M}_a \).
Using the nested linear codes \( \{ \mathcal{C}_i \mid  i =1,\ldots, a \} \) with the basis $\{\mathbf{b}_1,\ldots,\mathbf{b}_a\},$ the Construction-D lattice $\Lambda$ is defined as
\[
\Lambda = \left\{ \sum_{i=1}^a \sum_{j=1}^{k_i} p^{i-1} \beta_{ij} \tilde{\sigma}(b_j) \mid \beta_{ij} \in \{0, \ldots, p-1\} \right\} + p^a \mathbb{Z}^N,
\]
where \( \tilde{\sigma} \) is the natural embedding map from \( (\mathbb{Z}/\langle p\rangle)^N \) to \( \{0, \ldots, p-1\}^N \).
By using this hierarchical structure, Construction-D enables the formation of lattices with desirable properties such as good packing density and error correction capabilities, suitable for applications in communications and cryptography.
\subsection{Algebraic Construction-D  Lattices}
Let $ \theta$ be an algebraic integer such that $|\theta|^2 = p$, where $p$ is a natural prime number. Let $ K = \mathbb{Q}(\theta) $ be a number field and $\mathcal{O}_K = \mathbb{Z}[\theta]$ represent its ring of integers. This scenario is not applicable to all number fields; however, for a particular class of number fields known as monogenic number fields, there exists such element  that  generates both $ K $ and  $ \mathcal{O}_K$ \cite{Hassan}. 

Let \( p \mathcal{O}_K = \prod_{t=1}^{L} \mathfrak{P}_t \), where \( \mathfrak{P}_t \)'s are prime ideals in \( \mathcal{O}_K \), for $t=1,\ldots,L$. We assume that among these prime ideals,
there exists an index \( j_0 \) such that \( \mathfrak{P}_{j_0} \) is a principal ideal generated by an element \( \alpha_{j_0} \). Then, for all \( i \), we have \( \mathfrak{P}_{j_0}^i = \langle \alpha_{j_0}^i \rangle \). 
We further assume that the algebraic norm of \( \mathfrak{P}_{j_0} \) is \( N(\mathfrak{P}_{j_0}) = p \), which implies that the inertia degree of \( \mathfrak{P}_{j_0} \), denoted by \( f_{j_0} \), is one.
With this assumption, we have \( \left[ \frac{\mathcal{O}_K}{\mathfrak{P}_{j_0}} : \frac{\mathbb{Z}}{\langle p\rangle} \right] = 1 \). For \( i \in \{1, \ldots, a\} \), we define the ring homomorphism \( \phi_i : \mathcal{O}_K \to \frac{ \mathfrak{P}_{j_0}^{i-1}
}{\mathfrak{P}_{j_0}^i} \) as follows:
\begin{eqnarray}\label{phi_def}
  \phi_i(x) = \left[  x - x \pmod{\mathfrak{P}_{j_0}^{i-1}} \right]\pmod{\mathfrak{P}_{j_0}^i} .
\end{eqnarray}
To see why \( \phi_i \) is well-defined, we consider an element \( x \in \mathcal{O}_K \). Then \( \hat{x} = x - x \pmod{\mathfrak{P}_{j_0}^{i-1}} \) is an element in \( \mathfrak{P}_{j_0}^{i-1} \). Hence, \( \hat{x} = \alpha_{j_0}^{i-1} r \) for some \( r \in \mathcal{O}_K \). The map \( \phi \) then returns an element in $\mathfrak{P}_{j_0}^{i-1}/\mathfrak{P}_{j_0}^i$. The following lemma will be useful in the subsequent discussion.

Let $M$ be an $R$-module, where $R$ is a ring, and let $I$ be an ideal of $R$. The set $IM = \{\sum_{i=1}^k a_i x_i \mid k \in \mathbb{N}, a_i \in I, x_i \in M\}$ is a submodule of $M$. The quotient module $M/IM$ is an $R/I$-module with the action defined by:
\[ (r + I)(x + IM) = rx + IM, \]
for all $r \in R$ and $x \in M$. This action is well-defined.
\begin{lemma}[\emph{\cite{sahai2003algebra}, Lemma 4.4.15}]\label{module}
Let $R$ be a ring, $I$ a proper ideal of $R$, and $M$ a free $R$-module with basis $B$. Then $M/IM$ is a free $R/I$-module with basis $\pi(B)$, where $\pi: M \to M/IM$ is the canonical epimorphism, and $|B| = |\pi(B)|$.
\end{lemma}

Let $R = \mathcal{O}_K$ and $M = I = \mathfrak{P}_{j_0} = \alpha_{j_0}\mathcal{O}_K$. Adopting the notation of Lemma \ref{module}, it follows that $IM = \mathfrak{P}_{j_0}^2$.  The quotient $\mathfrak{P}_{j_0}/\mathfrak{P}_{j_0}^2$ is an $\mathcal{O}_K/\mathfrak{P}_{j_0}$-module, and thus an $\mathbb{F}_p$-vector space, where $\mathbb{F}_p \cong \mathcal{O}_K/\mathfrak{P}_{j_0}$. 
Define the map $\psi: \mathcal{O}_K/\alpha_{j_0}\mathcal{O}_K \to \alpha_{j_0}\mathcal{O}_K/\alpha_{j_0}^2\mathcal{O}_K$ by $\psi(r + \alpha_{j_0}\mathcal{O}_K) = \alpha_{j_0}r + \alpha_{j_0}^2\mathcal{O}_K$.  It is readily verified that $\psi$ is an isomorphism of $(\mathcal{O}_K/\alpha_{j_0}\mathcal{O}_K)$-modules.
Consequently, we have the isomorphism $\mathcal{O}_K/\mathfrak{P}_{j_0} \cong \mathfrak{P}_{j_0}/\mathfrak{P}_{j_0}^2 \cong \mathbb{F}_p$ of vector spaces.  By induction on $i$, setting $M = \alpha_{j_0}^i \mathcal{O}_K$, $R = \mathcal{O}_K$, and $I = \alpha_{j_0} \mathcal{O}_K$, we establish that $\mathfrak{P}_{j_0}^{i-1}/\mathfrak{P}_{j_0}^i \cong \mathbb{F}_p$ as vector spaces, for all $i$.
\begin{remark}
To maintain consistency with the field $\mathbb{F}_p$, we utilize the isomorphisms established earlier to map the output of $\phi_i$ to elements of $\mathbb{F}_p$. For each $i$, let $\psi_i: \mathfrak{P}{j_0}^{i-1}/\mathfrak{P}{j_0}^i \to \mathbb{F}_p$ be such an isomorphism, thus simplifying the definition of Construction-D lattices.

\end{remark}

Let \(  \mathcal{C}_1 \subset \cdots \subset  \mathcal{C}_a \) be a sequence of nested linear codes over \( \mathbb{F}_p \), where each \(  \mathcal{C}_i \) has parameters \( [N, k_i] \) for \( i = 1, \ldots, a \). There exists a basis \( \{ \mathbf{m}_1, \ldots, \mathbf{m}_N \} \) for the vector space \( \mathbb{F}_p^N \) such that \( \left\{\mathbf{m}_1, \ldots, \mathbf{m}_{k_i}\right\} \) spans \( \mathcal{C}_i \) and provides the generator matrix \( \mathbf{M}_i \) for \( \mathcal{C}_i \), where \( i = 1, \ldots, a \).
Now, we define our Construction-D lattice using the following steps:
\begin{enumerate}
	\item Define nested linear codes $\mathcal{C}_i=\left\{\mathbf{x}=\mathbf{w}_i\mathbf{M}_i|\mathbf{w}_i\in\mathbb{F}_p^{k_i}\right\}$ for $i=1,\ldots,a,$
	\item  We use the  symbol \( \Phi_i \) to denote the component-wise expansion of $\phi_i$ in \eqref{phi_def}  as  \( \Phi_i: \mathcal{O}_K^N \to \left( \frac{\mathfrak{P}_{j_0}^{i-1}}{\mathfrak{P}_{j_0}^i} \right)^N \) such that 
	\begin{equation}\label{Phi_def}
		\Phi_i(\mathbf{x})=(\phi_i(x_1),\dots,\phi_i(x_N)),
	\end{equation}
	for all $\mathbf{x}\in\mathcal{O}_K^N$. In a similar way, we consider $\Psi_i$ to denote the component-wise expansion of $\psi_i$. Now, we define the Construction-D lattice $\Lambda$ as follows: 
	 \begin{equation}\label{ConstructionD}
	 	\Lambda=\left\{\mathbf{x}\in\mathcal{O}_K^N\,\mid \, \Psi_i\left(\Phi_i(\mathbf{x})\right)\in \mathcal{C}_i,\,\, \text{for}\ i=1,\ldots,a\right\}.
	 \end{equation}
\end{enumerate} 
\begin{remark}
In Section~\ref{gen_mat_sec}, we derive a semi-systematic generator matrix for Construction-D lattices under the assumptions and notation established in this section.  Initially, we assumed $\mathfrak{P}_{j_0}$ to be a principal prime ideal with norm $p$, facilitating a connection between the generator matrices of the underlying codes and the generator matrix of our Construction-D lattice.  However, this assumption proves unnecessary.  Indeed, our proposed decoding algorithm operates directly on the lattice generator matrix, independent of the underlying codes.  Therefore, knowledge of the connection to the underlying codes is not required.  Investigating decoders that leverage this additional assumption remains a promising area for future research.
\end{remark}
\subsection{Generator Matrix of Algebraic Construction-D Lattices}\label{gen_mat_sec}
In the study of lattice coding, the generator matrix plays a crucial role in defining the structure and properties of a lattice. For Construction-D algebraic lattices, the generator matrix is built by leveraging the generator matrices of a sequence of nested linear codes, the ring of integers of an underlying number field, and a set of prime ideals. This construction method allows us to create lattices that are well-suited for applications such as error correction in wireless communication, physical layer security, and cryptography. This section provides a detailed exploration of the construction process of the generator matrix for Construction-D lattices. We analyze how the matrices from the underlying codes are combined with the matrices derived from the canonical embedding of the ring of integers and the prime ideals.
	
Let \( p \) be a prime number, and consider a chain of nested linear codes \( \mathcal{C}_1 \subset \mathcal{C}_2 \subset \cdots \subset \mathcal{C}_a \), over \( \mathbb{F}_p \), with parameters \([N, k_i]\) for $i=1,\ldots,a$. Let \( \mathbf{M} \) be the generator matrix of the linear code \( \mathcal{C}_a \). This matrix can be expressed in a structured form that reflects the inclusion relationships between the codes. Specifically, we can represent \( \mathbf{M} \) as:
\begin{IEEEeqnarray}{rCl}
&&\mathbf{M} = \\
&&\left[\begin{array}{cccccc}
\mathbf{I}_{k_1} & \mathbf{0}_{k_1\times\partial k_2} & \mathbf{0}_{k_1\times\partial k_3} & \cdots & \mathbf{0}_{k_1\times\partial k_a} & \mathbf{B}_1 \\
\mathbf{0}_{\partial k_2\times k_1} & \mathbf{I}_{\partial k_2} & \mathbf{0}_{\partial k_2\times\partial k_3} & \cdots & \mathbf{0}_{\partial k_2\times\partial k_a} & \mathbf{B}_2 \\
\mathbf{0}_{\partial k_3\times k_1} & \mathbf{0}_{\partial k_3\times \partial k_2 } & \mathbf{I}_{\partial k_3} &  \cdots & \mathbf{0}_{\partial k_3\times\partial k_a}  & \mathbf{B}_3 \\
\vdots & \vdots & \vdots & \ddots & \vdots & \vdots \\
\mathbf{0}_{\partial k_a \times k_1} & \mathbf{0}_{\partial k_a\times \partial k_2}  & \mathbf{0}_{\partial k_a\times \partial k_3}& \cdots  & \mathbf{I}_{\partial k_a} & \mathbf{B}_a
\end{array} \right] , \nonumber
\end{IEEEeqnarray}
where $\partial k_i \triangleq k_i-k_{i-1}$, and each \( \mathbf{B}_i \) is a matrix of size \( \partial k_i \times (N - k_a)\) for \( i = 1, \dots, a \), with \( k_0 = 0 \). The first \( k_1 \) rows of this matrix generate the code \( \mathcal{C}_1 \), the first \( k_2 \) rows generate the code \( \mathcal{C}_2 \), and so on, up to the code \( \mathcal{C}_a \).
	
For the number field \(K\) of degree \( n \), there are \( n \) distinct embeddings \( \sigma_t: K \to \mathbb{C} \), and the ring of integers \( \mathcal{O}_K \) has rank \( n \). Let \( \{w_1, w_2, \dots, w_n\} \) be a \( \mathbb{Z} \)-basis for \( \mathcal{O}_K \); the generator matrix for the lattice formed by \( \sigma(\mathcal{O}_K) \) is then given by:
\begin{equation}
\mathbf{T}_1 = \left[\begin{array}{cccc}
\sigma_1(w_1) & \sigma_2(w_1) & \cdots & \sigma_n(w_1) \\
\vdots & \vdots & \ddots & \vdots \\
\sigma_1(w_n) & \sigma_2(w_n) & \cdots & \sigma_n(w_n) 
\end{array} \right] .
\end{equation}
Let $\left\{\mu_1,\ldots,\mu_n\right\}$ be the \( \mathbb{Z} \)-basis of \( \mathfrak{P}_{j_0} \), where each \( \mu_i = \sum_{j=1}^{n} \mu_{i,j} w_j \) for \( i = 1, \dots, n \). Let \( \mathbf{T}_2 \) be a generator matrix for the ideal \( \sigma(\mathfrak{P}_{j_0}) \), given by:	
\begin{IEEEeqnarray}{rCl}
\mathbf{T}_2 &=& \left[\begin{array}{cccc}
\sigma_1(\mu_1) & \sigma_2(\mu_1) & \cdots & \sigma_n(\mu_1) \\
\vdots & \vdots & \ddots &\vdots\\
\sigma_1(\mu_n) & \sigma_2(\mu_n) & \cdots & \sigma_n(\mu_n)
\end{array} \right] \nonumber\\
&=& \left[ \begin{array}{cccc}
t_{1,1} & t_{1,2} &  \cdots & t_{1,n} \\
t_{2,1} & t_{2,2}&\cdots & t_{2,n} \\
\vdots & \vdots & \ddots & \vdots \\
t_{n,1}  &t_{n,2}& \cdots & t_{n,n}
\end{array} \right],
\end{IEEEeqnarray}
where $t_{i,j}=\sum_{j=1}^{n} \mu_{i,j} \sigma_j(w_j)$ for $i,j=1,\ldots,n$. 
The generator matrix for the Construction-D lattice \( \Lambda \) is then given in \eqref{generator_matrix}, where \( \otimes \) denotes the Kronecker product, and \( \mathbf{T}_i \)'s are the generator matrices for the ideals \( \sigma(\mathfrak{P}_{j_{0}}^i) \) for \( i = 2, \dots, a+1 \). Indeed, \( \mathbf{T}_i \) is derived in the same manner as we calculated \( \mathbf{T}_1 \) and \( \mathbf{T}_2 \), by considering the \(\mathbb{Z}\)-basis of \(\mathfrak{P}_{j_{0}}^i\). It is evident that the matrix \( \mathbf{G}_{\Lambda} \) contains exactly \( nN \) rows and columns.

\begin{figure*}[!t]
	\centering
		\begin{eqnarray}
			\mathbf{G}_{\Lambda} =
            \begin{bNiceMatrix}[margin]
				\mathbf{I}_{k_1} \otimes \mathbf{T}_1 & \mathbf{0}_{k_1\times\partial k_2} \otimes \mathbf{T}_1 & \cdots &\mathbf{0}_{k_1\times\partial k_a}\otimes\mathbf{T}_1 &  \mathbf{B}_1 \otimes \mathbf{T}_1 \\ 
				\mathbf{0}_{\partial k_2\times k_1}\otimes\mathbf{T}_2 & \mathbf{I}_{\partial k_2}\otimes\mathbf{T}_2 & \cdots & \mathbf{0}_{\partial k_2\times\partial k_a}\otimes\mathbf{T}_2 &  \mathbf{B}_2\otimes\mathbf{T}_2 \\ 
				\mathbf{0}_{\partial k_3\times k_1}\otimes\mathbf{T}_3 & \mathbf{0}_{\partial k_3\times \partial k_2 }\otimes\mathbf{T}_3 & \cdots & \mathbf{0}_{\partial k_3\times \partial k_a}\otimes\mathbf{T}_3 &  \mathbf{B}_3\otimes\mathbf{T}_3 \\ 
				\vdots & \vdots & \ddots & \vdots & \vdots \\ 
				\mathbf{0}_{\partial k_a \times k_1}\otimes\mathbf{T}_a & \mathbf{0}_{\partial k_a\times \partial k_2} \otimes\mathbf{T}_a &  \cdots  & \mathbf{I}_{\partial k_a}\otimes\mathbf{T}_a & \mathbf{B}_a\otimes\mathbf{T}_a \\ 
				\mathbf{0}_{\partial k_{a+1} \times k_1}\otimes\mathbf{T}_{a+1}& \mathbf{0}_{\partial k_{a+1}\times \partial k_2}\otimes\mathbf{T}_{a+1}  & \cdots & \mathbf{0}_{\partial k_{a+1}\times \partial k_a}\otimes\mathbf{T}_{a+1}  &  \mathbf{I}_{N-k_a}\otimes\mathbf{T}_{a+1}
                \end{bNiceMatrix}.
            \label{generator_matrix}
		\end{eqnarray}
	\caption{The generator matrix of the Construction-D lattice $\Lambda$.}
	\rule{\textwidth}{0.5pt}
    \vspace{-0.95cm}
\end{figure*}
\begin{proposition}\label{prop3}
Let $K$ be a number field with signature $(r_1,r_2)$ and $\mathfrak{P}_{j_0}=\alpha_{j_0}\mathcal{O}_K$ be a prime ideal with norm $N(\mathfrak{P}_{j_0})=p$. The Construction-D lattice $\Lambda$ based on number field $K$, defined in \eqref{ConstructionD}, is a sublattice of $\sigma(\mathcal{O}_K^N)$ with discriminant $ \Delta(\Lambda) 
       = \left(\frac{\sqrt{|\Delta_K}|}{2^{r_2}}\right)^{2N} p^{2aN - 2\sum_{i=1}^{a}k_i}$.
\end{proposition}
\begin{prof}
Without loss of generality, let us assume $a=2$, which means there are only two codes $\mathcal{C}_1$ and $\mathcal{C}_2$ with parameters $[k_1, N]$ and $[k_2, N]$, respectively. Then, generalization to all values of $a$ is straightforward. Consider the integer vectors $\mathbf{u}_1 = (\mathbf{u}_{1,1}, \dots, \mathbf{u}_{1,k_1})$, $\mathbf{u}_2 = (\mathbf{u}_{2,1}, \dots, \mathbf{u}_{2,(k_2-k_1)})$, and $\mathbf{u}_3 = (\mathbf{u}_{3,1}, \dots, \mathbf{u}_{3,(N-k_2)})$, where each component belongs to $ \mathbb{Z}^n$.  Let $\mathbf{T}_1$, $\mathbf{T}_2$, and $\mathbf{T}_3$ be the generator matrices of the lattices obtained from the ring $\sigma(\mathcal{O}_K)$ and ideals $\sigma(\mathfrak{P}_{j_0})$ and $\sigma(\mathfrak{P}^2_{j_0})$, respectively. Then, we have
\begin{multline*}
	(\mathbf{u}_1, \mathbf{u}_2, \mathbf{u}_3) 
    \left[\begin{array}{ccc}
		\mathbf{I}_{k_1} \otimes \mathbf{T}_1 & \mathbf{0} \otimes \mathbf{T}_1 & \mathbf{B}_1 \otimes \mathbf{T}_1 \\
		\mathbf{0} \otimes \mathbf{T}_2 & \mathbf{I}_{k_2-k_1} \otimes \mathbf{T}_2 & \mathbf{B}_2 \otimes \mathbf{T}_2 \\
		\mathbf{0}\otimes\mathbf{T}_3 & \mathbf{0}\otimes\mathbf{T}_3 & \mathbf{I}_{N-k_2} \otimes \mathbf{T}_3 
	\end{array} \right] \\
	= 
	\Big[ 
	\mathbf{u}_1 (\mathbf{I}_{k_1} \otimes \mathbf{T}_1),
	\mathbf{u}_2 (\mathbf{I}_{k_2-k_1} \otimes \mathbf{T}_2), \\
	\mathbf{u}_1 (\mathbf{B}_1 \otimes \mathbf{T}_1) + 
	\mathbf{u}_2 (\mathbf{B}_2 \otimes \mathbf{T}_2) + 
	\mathbf{u}_3 (\mathbf{I}_{N-k_2} \otimes \mathbf{T}_3) 
	\Big] \\
	= 
	\Big[ \left(\mathbf{u}_{1,1}\mathbf{T}_1,\ldots,\mathbf{u}_{1,k_1}\mathbf{T}_1 \right),\left(\mathbf{u}_{2,1}\mathbf{T}_2,\ldots,\mathbf{u}_{2,(k_2-k_1)}\mathbf{T}_2 \right),
    \\
\mathbf{u}_1 (\mathbf{B}_1 \otimes \mathbf{T}_1)+\mathbf{u}_2 (\mathbf{B}_2 \otimes \mathbf{T}_2)+  \left(\mathbf{u}_{3,1}\mathbf{T}_3,\ldots,\mathbf{u}_{3,(N-k_2)}\mathbf{T}_3 \right) \Big].
\end{multline*}
Using the definition of $\mathbf{T}_1,\mathbf{T}_2,\mathbf{T}_3$, for $i=1,\ldots,k_1$, we have $\mathbf{u}_{1,i}\mathbf{T}_1\in \sigma(\mathcal{O}_K)$, and for $j=1,\ldots,\partial k_2$,  $\mathbf{u}_{2,j}\mathbf{T}_2\in \sigma(\mathfrak{P}_{j_0})$ and for $t=1,\ldots,N-k_2$, we have $\mathbf{u}_{3,t}\mathbf{T}_3\in \sigma(\mathfrak{P}_{j_0})^2$. Now consider $\mathbf{B}_l=(b_{i,j}^{l})$, for $l=1,2,3$, which gives us
\begin{multline*}
	(\mathbf{u}_1, \mathbf{u}_2, \mathbf{u}_3) \mathbf{G}_{\Lambda}=\\
	\Big[ 
	(\mathbf{u}_{1,1} \mathbf{T}_1, \ldots, \mathbf{u}_{1, k_1} \mathbf{T}_1), 
	(\mathbf{u}_{2,1} \mathbf{T}_2,  \ldots, \mathbf{u}_{2, (k_2-k_1)} \mathbf{T}_2), \\
	\Big( 
	\sum_{i=1}^{k_1} b^1_{i, 1} \mathbf{u}_{1, i} \mathbf{T}_1, \ldots, 
	\sum_{i=1}^{k_1} b^1_{i, (N-k_2)} \mathbf{u}_{1 ,i} \mathbf{T}_1 
	\Big) 
	+ \\
	\Big( 
	\sum_{i=1}^{k_2-k_1} b^2_{i ,1} \mathbf{u}_{2 ,i} \mathbf{T}_2, \ldots, 
	\sum_{i=1}^{k_2-k_1} b^2_{i ,(N-k_2)} \mathbf{u}_{2, i} \mathbf{T}_2 
	\Big) 
	+ \\
	(\mathbf{u}_{3,1} \mathbf{T}_3, \ldots, \mathbf{u}_{3, (N-k_2)} \mathbf{T}_3) 
	\Big].
\end{multline*}
We apply $\sigma^{-1}$ on the resulting vector and denote the obtained vector by $\mathbf{x}=(\mathbf{x}_1,\mathbf{x}_2,\mathbf{x}_3)\in\mathcal{O}_K^{k_1}\times\mathfrak{P}_{j_0}^{k_2-k_1}\times\mathcal{O}_K^{N-k_2}$ which has the following form:
\begin{eqnarray*}
    \mathbf{x}&=&\Bigg( (x_{1,1},\ldots,x_{1,k_1}),(x_{2,1},\ldots,x_{2,k_2-k_1}),\\
   && \Big(\sum_{i=1}^{k_1}b_{i,1}^1x_{1,i},\ldots,\sum_{i=1}^{k_1}b_{i,(N-k_2)}^1x_{1,i}\Big)+\\
  &&      \Big(\sum_{i=1}^{k_2-k_1}b_{i,1}^2x_{2,i},\ldots,\sum_{i=1}^{k_2-k_1}b_{i,(N-k_2)}^2x_{2,i}\Big)+\\
  &&\qquad\qquad\qquad\qquad\quad(x_{3,1},\ldots,x_{3,N-k_2})\Bigg).
\end{eqnarray*}
Since the first \( k_2 \) components of \( \mathbf{x} \) are arbitrary elements of \( \mathcal{O}_K \) and \( \mathcal{O}_K/\mathfrak{P}_{j_0} \cong \mathbb{F}_p \), applying the homomorphism \( \Phi_1 \) to \( \mathbf{x} \), which computes modulo \( \mathfrak{P}_{j_0} \) component-wise, yields \( \bar{\mathbf{x}} = \Phi_1(\mathbf{x})\in\mathcal{C}_1 \) in the following form:
\begin{eqnarray*}
    \bar{\mathbf{x}}&=&\bigg( (\bar{x}_{1,1},\ldots,\bar{x}_{1,k_1}),(0\ldots,0),\\
   && \Big(\sum_{i=1}^{k_1}b_{i,1}^1\bar{x}_{1,i},\ldots,\sum_{i=1}^{k_1}b_{i,(N-k_2)}^1\bar{x}_{1,i}\Big)\bigg)\\
   &=&(\bar{x}_{1,1},\ldots,\bar{x}_{1,k_1}) \left[
  \begin{array}{ccc}
    \mathbf{I}_{k_1} & \mathbf{0}_{(k_2-k_1)} & \mathbf{B}_1 \\
  \end{array}
\right],
\end{eqnarray*}
where the last two summands in the third component of \( \mathbf{x} \) diminish when applying \( \Phi_1 \) because their components belong to \( \mathfrak{P}_{j_0} \) and \( \mathfrak{P}_{j_0}^2 \), which is a subset of \( \mathfrak{P}_{j_0} \). It is evident that $\Psi_1$ behaves as the identity map and satisfies $\Psi_1\left(\Phi_1(\mathbf{x})\right) \in \mathcal{C}_1$.
By applying $\Phi_2$, we obtain the vector $\bar{\bar{\mathbf{x}}} = (\mathbf{x} - \bar{\mathbf{x}}) \pmod{\mathfrak{P}_{j_0}^2}$, which belongs to $\left(\mathfrak{P}_{j_0}/\mathfrak{P}_{j_0}^2\right)^N$. Consequently, applying $\Psi_2$ maps this vector to an element in $\mathbb{F}_p^N$. Furthermore, 
\vspace{-0.5cm}
\begin{eqnarray*}
		\bar{\bar{\mathbf{x}}} &=& 
        \Bigg( (\bar{\bar{x}}_{1,1},\ldots,\bar{\bar{x}}_{1,k_1}),(\bar{\bar{x}}_{2,1},\ldots,\bar{\bar{x}}_{2,k_2-k_1}),\\
   && \Big(\sum_{i=1}^{k_1}b_{i,1}^1\bar{\bar{x}}_{1,i},\ldots,\sum_{i=1}^{k_1}b_{i,(N-k_2)}^1\bar{\bar{x}}_{1,i}\Big)+\\
  &&      \Big(\sum_{i=1}^{k_2-k_1}b_{i,1}^2\bar{\bar{x}}_{2,i},\ldots,\sum_{i=1}^{k_2-k_1}b_{i,(N-k_2)}^2\bar{\bar{x}}_{2,i}\Big)\Bigg)\\
  &=& (\bar{\bar{\mathbf{x}}}_{1},\bar{\bar{\mathbf{x}}}_{2})
\left[
  \begin{array}{ccc}
			\mathbf{I}_{k_1} & \mathbf{0} & \mathbf{B}_1 \\
			\mathbf{0} & \mathbf{I}_{k_2-k_1} & \mathbf{B}_2 
	\end{array}
    \right],
\end{eqnarray*}
for which we have $\Psi_2(\Phi_2(\mathbf{x})) = \Psi_2(\bar{\bar{\mathbf{x}}}) \in \mathcal{C}_2$, as needed.
As we know, $|\det(\mathbf{T}_1)| =\mathrm{vol}(\sigma(\mathcal{O}_K))= 2^{-r_2} \sqrt{|\Delta_K|}$, therefore, the absolute value of the determinant of $\mathbf{G}_{\Lambda}$ can be computed as	
\begin{equation*}\label{volume}
\begin{aligned}
    |\det(\mathbf{G}_{\Lambda})| &= |\det(\mathbf{I}_{k_1} \otimes \mathbf{T}_1)| \prod_{i=2}^{a} |\det(\mathbf{I}_{k_{i}-k_{i-1}} \otimes \mathbf{T}_i)| \\
    &\quad \times |\det(\mathbf{I}_{N-k_a} \otimes \mathbf{T}_{a+1})| \\
    &= |\det(\mathbf{T}_1)|^{k_1} \prod_{i=2}^{a} |\det(\mathbf{T}_i)|^{k_{i}-k_{i-1}} \\ 
    &\quad \times|\det(\mathbf{T}_{a+1})|^{N-k_a} \\
    &= \left( \frac{\sqrt{|\Delta_K|}}{2^{r_2}} \right)^N N(\mathfrak{P}_{j_0})^{aN - \sum_{i=1}^{a} k_i} \\
    &= \left( \frac{\sqrt{|\Delta_K|}}{2^{r_2}} \right)^N p^{aN - \sum_{i=1}^{a} k_i},
\end{aligned}
\vspace{-0.2cm}
\end{equation*}
where $\det(\mathbf{T}_i)=\mathrm{vol}(\sigma(\mathfrak{P}_{j_0}^{i-1}))=\Big( \frac{\sqrt{|\Delta_K|}}{2^{r_2}} \Big)p^{i-1}$ for $i=2,\dots,a+1$. 
On the other hand, since $\Lambda \subset \sigma(\mathcal{O}_K)^N$, let us consider $\mathcal{L}(\sigma(\mathcal{O}_K)^N, \Lambda) = \sigma(\mathcal{O}_K)^N \cap \mathcal{V}(\Lambda)$. Similar to \cite[Theorem 4]{strey2017lattices}, we can demonstrate that the elements of the following set $C$ are linearly independent, 
$$C=\left\{\sum_{s=0}^{a-1}\sum_{i=k_s+1}^{k_{s+1}}\beta_{i}^{(s+1)}\Phi_{s+1}^{-1}(\Psi_{s+1}^{-1}(\mathbf{b}_i))) \,\mid\beta_{i}^{(s+1)}\in\mathbb{Z}_{p^{a-s}} \right\},$$
where $k_0 = 0$, and $\{\mathbf{b}_1, \ldots, \mathbf{b}_{k_a}\}$ is the basis of $\mathcal{C}_a$. Hence, 
$$|C|=\prod_{s=0}^{a-1}\left(p^{(a-s)}\right)^{(k_{s+1}-k_s)}=p^{\left(\sum_{i=1}^{a}k_i\right)},$$
and the number of cosets is $|\mathcal{L}(\sigma(\mathcal{O}_K)^N,\Lambda)|=(p^{a})^N/|C|=p^{aN-\sum_{i=1}^a k_i}$.
Additionally, we have 
$|\mathcal{L}(\sigma(\mathcal{O}_K)^N,\Lambda)|=\dfrac{|\sigma(\mathcal{O}_K)^N|}{|\Lambda|}=\dfrac{\mathrm{vol}(\Lambda)}{\mathrm{vol}(\sigma(\mathcal{O}_K)^N)}.$
Hence, $\mathrm{vol}(\Lambda)=p^{aN-\sum_{i=1}^{a}k_i}\mathrm{vol}(\sigma(\mathcal{O}_K)^N)=\left(\dfrac{\sqrt{|\Delta_K|}}{2^{r_2}}\right)^Np^{aN-\sum_{i=1}^{a}k_i}.$ 
This completes the proof, as we have shown that the lattice generated by $\mathbf{G}_{\Lambda}$ forms a sublattice of the Construction-D lattice, based on the given family of underlying codes, with the same volume. Therefore, it must be equal to the entire lattice.
\hfill
\end{prof}
\begin{example}
Let $p=5$, and consider the cyclotomic field $K=\mathbb{Q}(\xi_5)$ with the totally real maximal subfield $K^+=\mathbb{Q}(\theta)$, where $\theta=\xi_5+\xi_5^{-1}$. Hence, the ring of integers of $K^+$ is $\mathcal{O}_{K^+}=\mathbb{Z}[\theta]$. The minimal polynomial of $\theta$ is $f(x)=x^2+x-1$ and $[K^+:\mathbb{Q}]=\dfrac{p-1}{2}=2$. In this case, the canonical embeddings $\sigma_i:K^+\to\mathbb{C}$, for $i=1,2$, are
	\begin{IEEEeqnarray*}{rCl}
		\sigma_1(\theta)&=&\theta=\dfrac{\sqrt{5}-1}{2},\\
		\sigma_2(\theta)&=&\theta^2=\dfrac{-\sqrt{5}-1}{2}.
	\end{IEEEeqnarray*}
	Since the set $\{1,\theta\}$ forms a basis for the ring of integers, $\mathcal{O}_{K^+},$  the generator matrix $\mathbf{T}_1$ is
	\begin{equation}
		\mathbf{T}_1=\begin{bmatrix}
			1 & 1 \\
			\theta & \theta^2
		\end{bmatrix}.
	\end{equation}
	The discriminant of $K^+$ is $\Delta_{K^+}=|\det[\sigma_i(\theta_j)]|^2=5$. Considering the totally real number field $K^+$ gives us  the signature $(r_1,r_2)=(2,0)$ and full diversity $L=r_1=2$.
	
\noindent Now, the ideal $\langle p\rangle $ is completely ramified in $\mathcal{O}_{K^+}$ which means $5\mathcal{O}_{K^+}=\mathfrak{P}_{j_0}^2$, where the principal ideal $\mathfrak{P}_{j_0}$ is generated by $\alpha_{j_0}=2-\theta$. The norm of this ideal is $N(\mathfrak{P}_{j_0})=N(\alpha_{j_0})=5$ and one can easily prove that $\langle\alpha_{j_0}^i\rangle=\langle\alpha_{j_0}\rangle^i$. The bases for the ideals $\mathfrak{P}_{j_0}$ and $\mathfrak{P}_{j_0}^2$ are the sets $\{5,3+\theta\}$ and $\{5,5\theta\}$, respectively. Hence, the generator matrices for $\sigma(\mathfrak{P}_{j_0})$ and $\sigma(\mathfrak{P}_{j_0})^2$, which are denoted by $\mathbf{T}_2$ and $\mathbf{T}_3$, are 
	\begin{IEEEeqnarray}{rCl}
	    	\mathbf{T}_2&=&\begin{bmatrix}
			5 & 5 \\
			\dfrac{5+\sqrt{5}}{2}& \dfrac{5-\sqrt{5}}{2}
		\end{bmatrix}, \\
    \mathbf{T}_3&=&\begin{bmatrix}
			5 & 5 \\
			\dfrac{-5+5\sqrt{5}}{2}& \dfrac{-5-5\sqrt{5}}{2}
		\end{bmatrix},
	\end{IEEEeqnarray}
	with determinant $|\mathbf{T}_2|=-5\sqrt{5}$ and $|\mathbf{T}_3|=-25\sqrt{5}$, respectively.
	Finally, the homomorphisms $\Phi_i: \mathcal{O}_{K^+}^N \to \left( \frac{\mathfrak{P}_{j_0}^{i-1}}{\mathfrak{P}_{j_0}^i} \right)^N$, for $i=1,2$, are given by
	\begin{equation*}
		\Phi_i(\mathbf{x})=\big[\mathbf{x}-\mathbf{x}\pmod{\mathfrak{P}_{j_0}^{i-1}}\big] \pmod{\mathfrak{P}_{j_0}^i}.
	\end{equation*}
	For a set of nested linear codes $\mathcal{C}_1=\langle(100)\rangle$ and $\mathcal{C}_2=\langle(100),(011)\rangle$ over $\mathbb{F}_5$, the Construction-D lattice $\Lambda$ is defined as follows:
	\begin{equation*}
		\Lambda=\left\{\boldsymbol{\lambda}\in\mathcal{O}_{K^+}^3:\Phi_i(\boldsymbol{\lambda})\in \mathcal{C}_i,\ \,\mathrm{for}\,\, i=1, 2\right\},
	\end{equation*}
	with the generator matrix $\mathbf{G}_{\Lambda}$ as follows:
\begin{IEEEeqnarray*}{rCl}
	\begin{aligned}
		\mathbf{G}_{\Lambda} &= \begin{bmatrix}
			1 \otimes \mathbf{T}_1 & 0 \otimes \mathbf{T}_1 & 0 \otimes \mathbf{T}_1 \\
			0 \otimes \mathbf{T}_2 & 1 \otimes \mathbf{T}_2 & 1 \otimes \mathbf{T}_2 \\
			0 \otimes \mathbf{T}_3 & 0 \otimes \mathbf{T}_3 & 1 \otimes \mathbf{T}_3
		\end{bmatrix} 
		= \begin{bmatrix}
			\mathbf{T}_1 & \mathbf{0} & \mathbf{0} \\
			\mathbf{0} & \mathbf{T}_2 & \mathbf{T}_2\\
			\mathbf{0} & \mathbf{0}& \mathbf{T}_3 
		\end{bmatrix}.
	\end{aligned}
\end{IEEEeqnarray*}
\noindent Using $\mathbf{G}_{\Lambda}$, we obtain the volume of $\Lambda$ as $\mathrm{vol}(\Lambda)=|\mathbf{G}_{\Lambda}|=|\mathbf{T}_1||\mathbf{T}_2||\mathbf{T}_3|=|(-\sqrt{5})(-5\sqrt{5})(-25\sqrt{5})|=5^4\sqrt{5}$ which is equivalent to the value obtained using the formula provided in Proposition~\ref{prop3}.
\end{example}

In this section, we developed a semi-systematic generator matrix for Construction-D lattices by leveraging the algebraic properties of number fields and their ideals. The structured nature of the generator matrix allows for efficient representation, simplifies encoding, and enhances error resilience, while also connecting the lattice's fundamental volume to its algebraic properties. These features significantly improve the overall performance of the proposed Construction-D lattices in practical communication applications.
\section{Construction-D Lattices over Block Fading Channels}\label{bf}
In this section, we explore the key performance metrics  of algebraic lattices when utilized  in a BF channel. 
BF channels can be viewed as a specific case of flat fading channels, characterized by a constant channel gain over each block of symbols while allowing for independent variations across different blocks. This model is essential for evaluating the performance of communication systems where channel conditions change gradually compared to the symbol duration.
\subsection{Channel Model}
The flat fading channel is a fundamental model in wireless communications, representing environments where all frequency components of the transmitted signal experience the same level of fading. This model is particularly relevant in scenarios where the bandwidth of the transmitted signal is much smaller than the coherence bandwidth of the channel, such as in narrowband wireless communication systems. 
\subsubsection{Mathematical Formulation}
Consider a flat fading channel where the transmitted symbols are affected by an $n\times n$ flat fading diagonal matrix $\mathbf{H}_f=\text{diag}(\mathbf{h})$ where $\mathbf{h}=(h_1,\dots,h_n)$ and additive Gaussian noise. The received signal for $i=1,\dots,N$ can be expressed as:
\begin{equation}
	\mathbf{y}_i^t = \mathbf{H}_f \mathbf{x}_i^t + \mathbf{n}_i,
\end{equation}
where $\mathbf{y}_i\in\mathbb{R}^n$ is the received $n$ dimensional signal, $\mathbf{x}_i\in\mathbb{R}^n$ is the transmitted $n$ dimensional signal, and each \( \mathbf{n}_i\in\mathbb{R}^n\) is the AWGN noise vector with Gaussian distribution \(\mathcal{N}(0, \sigma^2_{\mathcal{N}})\).
Each fading coefficient \( h_i \)  is modeled as a complex Gaussian random variable $h$ with its magnitude following a Rayleigh distribution.
In this case, the magnitude \( |h| \) of the fading coefficient follows a Rayleigh distribution:
\begin{equation}
	f_{|h|}(x) = \frac{x}{\sigma_h^2} e^{-x^2 / (2\sigma_h^2)}, \quad x \geq 0,
\end{equation}
where \( \sigma_h^2 \)  represents the variance of the fading process.
 Then, the received vector \( \mathbf{y} \) from a Rayleigh block fading channel with \( n \) fading blocks and coherence time \( N \) can be expressed as follows:
\begin{equation}\label{y}
    \mathbf{y}^t = (\mathbf{I}_N \otimes \mathbf{H}_F) \mathbf{x}^t +\mathbf{n}^t,
\end{equation}
where \( \mathbf{H}_F = \text{diag}(|h_1|, \dots, |h_n|) \), and the fading coefficients \( h_i \)'s are complex Gaussian random variables with variance \( \sigma_b^2 \), making \( |h_i| \)'s Rayleigh distributed with parameter \( \sigma_b^2 \) for  \( i = 1, \dots, n \). The noise vector \( \mathbf{n} = (\mathbf{n}_1, \ldots, \mathbf{n}_N) = (\nu_1, \dots, \nu_{nN}) \) consists of Gaussian noise samples \( \nu_i \sim N(0, \sigma_{\mathcal{N}}^2) \), for \( i = 1, \dots, nN \) \cite{Hassan}.
\subsubsection{Signal to Noise Ratio}
Let \( \mathbf{x}=(\mathbf{x}_1,\ldots,\mathbf{x}_N) \in \mathbb{R}^{nN} \) represent a frame consisting of \( N \) modulation symbols \( \mathbf{x}_i \), each with dimension \( n \). In this work, \( \mathbf{x} \) is selected from a Construction-D lattice \( \Lambda \) defined over a number field \( K \) of degree \( n \), with underlying nested \([N, k_i]\)-linear codes \( \mathcal{C}_i\), for $i=1,\dots,a$. This setup models communication over a BF channel with a fading block length of \( N \) \cite{Hassan}.
The instantaneous signal to noise ratio (SNR) for an infinite lattice constellation $\Lambda$ is given by:
\begin{IEEEeqnarray}{rCl}
	\gamma = \frac{\text{vol}(\Lambda)^{\frac{2}{nN}}}{\sigma^2_{\mathcal{N}}}.
\end{IEEEeqnarray}






\subsection{Error Performance and Characteristics of Good Lattices over BF Channels}
The error performance of lattices in BF channels is influenced by variations in channel conditions across different blocks. Understanding the relationship between BF characteristics, lattice structure, and decoding strategies is key to optimizing performance. This subsection examines lattices under varying fading conditions, focusing on the probability of error and its dependence on channel parameters. We benchmark our analysis using the maximum likelihood (ML) decoder for infinite lattices and compare the performance of Construction-D lattices with Construction-A lattices \cite{Hassan}.

In our communication model, the transmitted signal vector \( \mathbf{x} \) is drawn from an infinite lattice \( \Lambda \subset \mathbb{R}^{nN} \), represented as \( \Lambda = \{ \mathbf{u} \mathbf{G}_{\Lambda} \mid \mathbf{u} \in \mathbb{Z}^{nN} \} \), where \( \mathbf{G}_{\Lambda} \) is a full-rank generator matrix of size \( nN \times nN \). For each channel realization, the receiver observes a faded version of the lattice characterized by \( \mathbf{G}_{\Lambda_F} = (\mathbf{I}_N \otimes \mathbf{H}_F) \mathbf{G}_{\Lambda} \), with \( \mathbf{H}_F \) representing the fading coefficients and \( \mathbf{I}_N \) as the \( N \times N \) identity matrix. 
The transmitted lattice point \( \mathbf{x} \) consists of \( N \) modulation symbols \( \mathbf{x}_i \), each of dimension \( n \). We assume perfect channel state information (CSI) is available at the receiver, providing complete knowledge of the fading coefficients. Due to the geometric uniformity of lattices, we have \( \mathcal{V}(\mathbf{x}, \mathbf{h}) = \mathcal{V}(\mathbf{w}, \mathbf{h}) = \mathcal{V}_{\Lambda}(\mathbf{h}) \) for all \( \mathbf{x}, \mathbf{w} \in \Lambda \), allowing us to assume the transmission of the all-zero lattice vector. 
For a given fading realization \( \mathbf{h} \), the channel transition probabilities for an ML decoder, when the lattice point \( \mathbf{x} \in \Lambda \) is transmitted over a BF channel with an additive noise variance of \( \sigma^2_{\mathcal{N}} \) per dimension, can be expressed as follows:
\begin{multline*}
	\text{P}_e(\Lambda,\sigma^2_{\mathcal{N}}) =\mathbb{E}\left(\text{P}_e(\Lambda,\sigma^2_{\mathcal{N}}\mid \mathbf{h})\right)\\
	=1-\mathbb{E}\left(\int_{\mathcal{V}_{\Lambda}(\mathbf{h})} \left( \frac{1}{2 \pi \sigma^2_{\mathcal{N}}} \right)^{\frac{nN}{2}} \exp \left( -\frac{\|\mathbf{y} - \mathbf{h} \mathbf{x}\|^2}{2 \sigma^2_{\mathcal{N}}} \right)\right).
\end{multline*}
For a fixed
lattice $\Lambda$, the decoding error probability $\text{P}_e(\Lambda,\sigma^2_{\mathcal{N}})$ is clearly
a function of the SNR $\gamma$. In the rest of this paper, we denote it by $\text{P}_e(\gamma)$ when there is no potential for ambiguity. 
\begin{definition}[Diversity Order] The diversity order is defined as the asymptotic
	(for large SNR) slope of \( P_e(\gamma) \) in a $\log-\log$ scale, that is
\begin{equation}
	d = -\lim_{\gamma \to \infty} \frac{\log P_e(\gamma)}{\log(\gamma)}.
\end{equation}
\end{definition}
The diversity order, or diversity gain, measures a communication system's effectiveness in mitigating fading in wireless channels. A higher diversity order correlates with a faster decrease in error probability as SNR increases. In the case of using lattices  over BF channels, the diversity order depends on the lattice structure and fading channel characteristics.
\begin{definition}
    In a BF channel with \( n \) independent fading coefficients per lattice point, a lattice \( \Lambda \) is deemed to achieve full diversity under ML decoding if its diversity order \( d \) equals the number of fading blocks, such that \( d = n \).
\end{definition}
In communication systems employing lattices over BF channels, the point error probability \(P_e\) plays a critical role in determining overall system performance. 
To enhance reliability and mitigate point error probability, we utilize the following criteria established by \cite{Hassan} for constructing good lattices over BF channels:

\begin{enumerate}
    \item \textbf{Maximizing the diversity order} which is defined as
   $ L = \min_{\mathbf{x} \neq \mathbf{w} \in \Lambda} d_H(\mathbf{x} , \mathbf{w})$,
 where $d_H$ denotes the Hamming distance.
    \item \textbf{Increasing the minimum \(L\)-product distance} which is defined as \(
    d_{p,\text{min}}^L =\min_{\mathbf{x} \neq \mathbf{w} \in \Lambda} d_p^l(\mathbf{x},\mathbf{w})
    \), where $d_p^l(\mathbf{x},\mathbf{w})$ is the $l$-product distance of $\mathbf{x}$ from $\mathbf{w}$ when these two points differ  in $l$ components, that is $d_p^l(\mathbf{x},\mathbf{w})=\prod_{i=1}^{l}|x_i-w_i|$.
    \item \textbf{Minimizing the product kissing number \(\tau_p\)} which  is defined as the total number of lattice points at the minimum \(L\)-product distance.
\end{enumerate}
Let \( N_{d_{p,\text{min}}}(\Lambda) = \frac{d_{p,\text{min}}^{L}(\Lambda)}{\text{vol}(\Lambda)}\) represent the normalized minimum product distance of a lattice $\Lambda$. As stated in \cite{7282492}, the relationship between the discriminant \(\Delta_K\) of a totally real number field \( K \) with degree \( n \), and the normalized minimum product distance of the associated lattice \( \sigma(\mathcal{O}_K) \), is given by:
\[
\text{vol}(\sigma(\mathcal{O}_K)) = |\Delta_K|, \quad \text{and} \quad N_{d_{p,\text{min}}}(\sigma(\mathcal{O}_K)) = \frac{1}{\sqrt{|\Delta_K|}}.
\]
This relationship underscores the importance of selecting number fields with smaller discriminants for optimal lattice configurations, thereby enhancing performance in communication scenarios influenced by BF channels.
\subsection{Poltyrev Threshold and Poltyrev Outage Limit}
In lattice-based communication over BF channels, understanding the fundamental limits of reliable transmission is crucial. Two key concepts that characterize these limits are  the Poltyrev threshold and  the Poltyrev outage limit (POL).
\subsubsection{Poltyrev Threshold}
In the context of BF channels, 
the Poltyrev threshold defines the upper limit of the noise variance that can be tolerated without compromising the error performance of the lattice code.
For a BF channel with a fixed instantaneous fading $\mathbf{h}=\left(h_1,h_2,\dots,h_n\right)$ per lattice point, the Poltyrev threshold is expressed as:
\begin{equation}
	\sigma^2_{\text{max}}(\mathbf{h}) = \frac{| \det(\mathbf{G}_{\Lambda}) |^{\frac{2}{nN}}\prod_{i=1}^{n} h_i^{\frac{2}{n}}} {(2 \pi e)},
\end{equation}
where \( \mathbf{G}_{\Lambda} \) is the generator matrix of the lattice \( \Lambda \).
For reliable communication to be possible, the actual noise variance \( \sigma^2_{\mathcal{N}} \) must satisfy:
\begin{equation}
	\sigma^2_{\mathcal{N}} < \sigma^2_{\text{max}}(\mathbf{h}).
\end{equation}

This inequality indicates that the noise level must stay below the Poltyrev threshold to enable successful decoding of the lattice  with a low error probability. The Poltyrev threshold is influenced by both the channel's fading conditions and the geometric properties of the lattice.
\subsubsection{Poltyrev Outage Limit (POL)}
The POL for lattices over BF channels is introduced using the Poltyrev threshold in \cite{POLlimit}. It has been established that the POL achieves full diversity \(L\) for a channel with \(L\) independent fading blocks, confirming its ability to realize full diversity. For a given instantaneous fading vector \(\mathbf{h} = (h_1, \ldots, h_n)\), the POL can be expressed as:
\begin{equation}
	\begin{aligned}
		P_{\text{out}}(\gamma) &= \Pr \left( \sigma_{\mathcal{N}}^2 > \frac{|\det(\mathbf{G}_{\Lambda}) |^{\frac{2}{nN}}\prod_{i=1}^{n} h_i^\frac{2}{n}}{(2 \pi e)} \right) \\
		&= \Pr \left( \prod_{i=1}^{n} h_i^2 < \frac{(2 \pi e)^n}{\gamma^n} \right).
	\end{aligned}
\end{equation}
An outage occurs when the condition $\prod_{i=1}^{n} h_i < \left(\frac{2 \pi e}{\gamma}\right)^{n/2}$ is met, indicating that the channel quality is insufficient to maintain reliable communication at the given SNR level $\gamma$. For a given lattice, the frame error rate (FER) after lattice decoding over a BF channel can be compared to $P_{\text{out}}(\gamma)$ to assess the SNR gap and confirm the diversity order.
\section{Decoding of Full-Diversity Construction-D lattices}\label{decoding}
In this section, we introduce a decoding method for Construction-D lattices. Our approach utilizes segmentation and a low-dimensional lattice quantizer in the initial stages, followed by successive cancellation decoding. This technique effectively harnesses the advantages of optimal low-dimensional quantization to simplify the high-dimensional decoding process while enhancing overall performance. Subsequently, quantization is reapplied to refine the decoding of the remaining segments, thereby ensuring full diversity within the proposed Construction-D lattices.
\begin{definition}
    A \emph{lattice quantizer} is a mapping \( Q_{\Lambda} : \mathbb{R}^n \rightarrow \Lambda \), where \( \Lambda \) is a lattice in \( \mathbb{R}^n \), such that for each input vector \( \mathbf{y} \in \mathbb{R}^n \), the quantizer outputs the closest lattice point \( Q_{\Lambda}(\mathbf{y}) \in \Lambda \) according to  Euclidean distance. For each $\mathbf{y}\in\mathbb{R}^n$, we also define $\mathbf{y}\pmod{\Lambda}\triangleq\mathbf{y}-Q_{\Lambda}(\mathbf{y})$. 
\end{definition}

Let $\mathbf{y}$ be the received vector from a Rayleigh BF channel with $n$ fading blocks and coherence time $N$, which is given in  \eqref{y}. To retrieve the transmitted codeword \(\mathbf{x}\), we initiate the process by estimating the  vector \(\hat{\mathbf{u}} = (\hat{\mathbf{u}}_1, \dots, \hat{\mathbf{u}}_N)\), where  \(\hat{\mathbf{u}}_j = (\hat{u}_{j,1}, \ldots, \hat{u}_{j,n})\), for $j=1,\dots, N$, corresponds to the $j^{th}$ segment of \(\mathbf{x}\). Based on the structure of the generator matrix \(\mathbf{G}_{\Lambda}\) defined in  \eqref{generator_matrix}, the first \(k_a\) sub-vectors of \(\hat{\mathbf{u}}\) like $\hat{\mathbf{u}}_j$, $j=1,\ldots,k_a$, can be recovered using the quantization function \(Q_{\Lambda'_j}\), with \(\hat{\mathbf{u}}_j = Q_{\Lambda'_j}(\mathbf{y}_j^t)\). In this context, \(\Lambda'_j\) represents the lattice associated with the generator matrix \(\mathbf{T}'_j = \mathbf{H}_F \mathbf{G}_{j,j}^t\), where \(\mathbf{G}_{j,j}\) is the \(n \times n\) sub-matrix of the generator matrix \(\mathbf{G}_{\Lambda}\) corresponding to the rows and columns indexed from \((j-1)n+1\) to \(jn\). Additionally, \(\mathbf{y}_j = \mathbf{y}((j-1)n+1:jn)\), for \(j = 1, \ldots, N\).

After retrieving \(\hat{\mathbf{u}}_1, \dots, \hat{\mathbf{u}}_{k_a}\) and performing successive cancellation, we apply the quantization function again on \(\hat{\mathbf{y}}' = \mathbf{y}_j - \sum_{t=1}^{k_a} \hat{\mathbf{u}}_t \mathbf{G}_{t,t} \mathbf{H}_F\) to recover \(\hat{\mathbf{u}}_{k_a+1}, \dots, \hat{\mathbf{u}}_N\) as shown in Algorithm~\ref{alg}.

\begin{algorithm}
\caption{Full-Diversity Construction-D Lattices Decoding}
\label{alg}
\begin{algorithmic}[1] 
\Procedure{DEC}{$\mathbf{y}, \mathbf{G}_{\Lambda}, \mathbf{H}_F = \text{diag}(|h_1|, \dots, |h_n|)$}
    \State $\hat{\mathbf{u}} \gets 0_{1 \times nN}$
    \State $\hat{\mathbf{x}} \gets 0_{1 \times nN}$
    \For{$j = 1 : N$}
        \State $\mathbf{y}_j \gets \mathbf{y}((j - 1) \cdot n + 1 : j \cdot n)$
        \State  $\mathbf{G}_{j,j} \gets \big(\mathbf{G}_{\Lambda}(i,k)\big)_{i,k\in \left\{(j - 1) \cdot n + 1 ,\ldots, j \cdot n\right\}}$
        \If{$j \leq k_a$}
            \State $\hat{\mathbf{u}}_{j} \gets \arg \min_{\mathbf{u}_j \in \mathbb{Z}^n} \| \mathbf{y}_{j} - \mathbf{u}_j \mathbf{G}_{j,j} \mathbf{H}_F \|^2$
                \State $\hat{\mathbf{u}}((j - 1) \cdot n + 1 : j \cdot n) \gets \hat{\mathbf{u}}_{j}$
                 \State $\hat{\mathbf{x}}((j - 1) \cdot n + 1 : j \cdot n) \gets \hat{\mathbf{u}}_{j}\mathbf{G}_{j,j}$
        \Else
            \State $\mathbf{y}_j \gets \mathbf{y}_j - \sum_{t=1}^{k_a} \hat{\mathbf{u}}_t \mathbf{G}_{t,t} \mathbf{H}_F$
            \State $\hat{\mathbf{u}}_{j} \gets \arg \min_{\mathbf{u}_j \in \mathbb{Z}^n} \| \mathbf{y}_{j} - \mathbf{u}_j \mathbf{G}_{j,j} \mathbf{H}_F \|^2$
                \State $\hat{\mathbf{u}}((j - 1) \cdot n + 1 : j \cdot n) \gets \hat{\mathbf{u}}_{j}$
                \State $\hat{\mathbf{x}}((j - 1) \cdot n + 1 : j \cdot n) \gets \hat{\mathbf{u}}_{j}\mathbf{G}_{j,j}$
        \EndIf    
    \EndFor
    \State \Return $\hat{\mathbf{u}}, \hat{\mathbf{x}}$
\EndProcedure
\end{algorithmic}
\end{algorithm}
\subsection{Decoding Analysis}
This section assesses the performance of the proposed decoding algorithm analytically, highlighting its reliability and efficiency in BF channels.

In scenarios where the lattices \( \Lambda \) and \( \sigma(\mathfrak{P}_{j_0}^i) \) for $i=1,\ldots,a$ are used for communication over a BF channel with \( n \) fading blocks, the SNR expressions can be written as:
\begin{IEEEeqnarray}{rCl}
\gamma_{\Lambda} &=& \frac{\text{vol}(\Lambda)^{\frac{2}{nN}}}{\sigma_{\mathcal{N}}^2} = \left( \Delta_K^{\frac{N}{2}}p^{aN-\sum_{i=1}^ak_i} \right)^{2/nN} \frac{1}{\sigma_{\mathcal{N}}^2},\\
  \gamma_{\mathfrak{P}_{j_0}^i} &=& \frac{\text{vol}(\sigma(\mathfrak{P}_{j_0}^i))^{2/n}}{\sigma_{\mathcal{N}}^2} = \left( p^i\sqrt{\Delta_K} \right)^{2/n} \frac{1}{\sigma_{\mathcal{N}}^2}.
\end{IEEEeqnarray}
When both cases achieve full diversity, their error probabilities are mainly governed by \( 1/\gamma_{\Lambda}^n \) and \( 1/\gamma_{\mathfrak{P}_{j_0}^i}^n \), respectively. However, these two expressions are interconnected; in fact, we have \( \gamma_{\Lambda} =\left(p^{a-(\frac{\sum_{i=1}^ak_i}{N})-i}\right)^{\frac{2}{n}} \gamma_{\mathfrak{P}_{j_0}^i}\), which leads to the conclusion \( O(1/\gamma_{\mathfrak{P}_{j_0}^i}) = O(1/\gamma_{\Lambda}) \). Indeed, at high SNRs (i.e., \( \sigma_{\mathcal{N}}^2 \rightarrow 0 \)), \( \gamma_{\Lambda} \), and \( \gamma_{\mathfrak{P}_{j_0}^i} \) are effectively equivalent, allowing them to be represented simply by \( \gamma \) throughout the proof of Theorem~\ref{MainThm}.
\begin{theorem}\label{MainThm}
    Let \( \mathcal{C}_1\subset \mathcal{C}_2\subset \cdots\subset\mathcal{C}_a \subset \mathbb{F}_p^N \) be the underlying nested $[N,k_i]$-linear codes of a Construction-D lattice \( \Lambda \) with diversity \( n \). Then, \( \Lambda \) achieves full diversity over a BF channel with \( n \) fading blocks using the proposed decoder  in Algorithm~\ref{alg}.
\end{theorem}
\begin{prof}
Since retrieving $\hat{\mathbf{u}}$  and $\hat{\mathbf{x}}$ are equivalent, we define the decoding problem as recovering $\hat{\mathbf{u}}$ to simplify the notation. 
To optimally decode the lattice \(\Lambda\), one must determine \(\arg\min_{\mathbf{u} \in \mathbb{Z}^{nN}} \|\mathbf{y}^t - \big( \mathbf{u}\mathbf{G}_{\Lambda} (\mathbf{I}_N \otimes \mathbf{H}_F)\big)^t\|^2\). This problem is a particular case of the closest vector problem (CVP) in dimension $nN$, which is generally considered NP-hard. For certain cases and specific types of lattices, CVP can be solved more efficiently. For low-dimensional lattices (i.e., dimensions 2 or 3), the problem can be solved in polynomial time. We have utilized these practical low-dimensional quantizers in dimension $n$ as a subroutine within our decoding algorithm.
According to Algorithm~\ref{alg}, the initial \(k_a\) segments of \(\hat{\mathbf{u}}\) are recovered through \(n\)-dimensional optimal decoding (which maintains the diversity order) across various lattices:
\begin{IEEEeqnarray}{rCl}\label{de}
(\hat{\mathbf{u}}_1,\ldots,\hat{\mathbf{u}}_{k_a}) = \bigoplus_{j=1}^{k_a} \arg\min_{\mathbf{u}_j \in \mathbb{Z}^n} \|\mathbf{y}_j^t - \mathbf{H}_F \mathbf{G}_{j,j}^t \mathbf{u}_j^t\|^2,
\end{IEEEeqnarray}
where \(\bigoplus\) denotes the concatenation of \(k_a\) vectors of length \(n\). In the next step, after updating the segments of \(\mathbf{y}\), the same approach is applied to decode the remaining \(N - k_a\) segments as follows:
\begin{equation}\label{rest}
(\hat{\mathbf{u}}_{k_a+1},\ldots,\hat{\mathbf{u}}_{N}) = \bigoplus_{j=k_a+1}^{N} \arg\min_{\mathbf{u}_j \in \mathbb{Z}^n} \|\mathbf{y'}_j^t - \mathbf{H}_F \mathbf{G}_{j,j}^t \mathbf{u}_j^t\|^2,
\end{equation}
where $\mathbf{y'}_j=\mathbf{y}_j - \sum_{t=1}^{k_a} \hat{\mathbf{u}}_t \mathbf{G}_{t,t} \mathbf{H}_F$, for $j=k_a+1,\ldots, N$. Therefore, instead of performing a full ML decoding in dimension \(nN\), our approach simplifies this by solving the smaller optimization problems in equations \eqref{de} and \eqref{rest}, resulting in \(N\) separate ML decodings in dimension \(n\). 
To illustrate its functionality, we define \(\mathbf{w}_j\) as the zero vector of length \(nN\) except for the \(j^{th}\) segment, which is specified as follows:
\begin{IEEEeqnarray*}{rCl}
    \mathbf{w}_j((j - 1) \cdot n + 1 : j \cdot n)=
\begin{cases} 
\mathbf{y}_j^t - \mathbf{H}_F \mathbf{G}_{j,j}^t \mathbf{u}_j^t, & \hspace{-0.2cm} 1\leq j\leq k_a, \\ 
\mathbf{y'}_j^t - \mathbf{H}_F \mathbf{G}_{j,j}^t \mathbf{u}_j^t, & \hspace{-0.2cm} \text{otherwise}.
\end{cases}
\end{IEEEeqnarray*}
Then, using Triangle inequality, we have 
\begin{IEEEeqnarray*}{rCl}
\|\mathbf{y}^t - \big( \mathbf{u}\mathbf{G}_{\Lambda} (\mathbf{I}_N \otimes \mathbf{H}_F)\big)^t\|^2=\left\|\sum_{j=1}^{N}\mathbf{w}_j\right\|^2\leq \sum_{j=1}^{N}\left\|\mathbf{w}_j\right\|^2\\
=\sum_{j=1}^{k_a}  \|\mathbf{y}_j^t - \mathbf{H}_F \mathbf{G}_{j,j}^t\|^2             +
\sum_{j=k_a+1}^{N} \|\mathbf{y'}_j^t -\mathbf{H}_F \mathbf{G}_{j,j}^t \mathbf{u}_j^t\|^2.
\end{IEEEeqnarray*}
Thus, our algorithm addresses the right-hand side of the previous inequality. It is evident that this results in a sub-optimal solution.
Since each ML decoding instance achieves a diversity of \(n\), the probability of error for each decoded segment is bounded by \(\gamma_{\mathfrak{P}_{j_0}^i}^{-n}\), for some $1\leq i\leq a$, which is equivalent to \(\gamma^{-n}\) based on our assumptions. An error in the concatenated result occurs if at least one of these individual decodings fails, giving an upper bound on the point error probability of our decoder as \(N \gamma^{-n}\), which confirms that the diversity order \(n\) is achieved. \hfill
\end{prof}
\subsection{Complexity Analysis of the Decoding Algorithm}
The proposed decoding algorithm utilizes a hybrid approach that integrates ML decoding, successive cancellation, and a subsequent ML decoding phase.

In the first phase of the decoding process, the algorithm applies ML decoding to the initial \( k_a \) components. The complexity associated with this phase is expressed as \( O(k_a \cdot f(n)) \), where \( f(n) \) denotes the complexity of the ML decoder in dimension \( n \). Since \( n \) represents  the small number of fading coefficients, it can be regarded as a polynomial in terms of \( n \), which is negligible compared to $k_a$ and we consider the complexity of this phase as linear with respect to $k_a$.
Following the initial ML decoding, the algorithm performs a successive cancellation method for the remaining \( N-k_a \) components. 
Since \( \mathbf{G}_{t,t} \) is an \( n \times n \) matrix, the vector-matrix multiplication \( \hat{u}_t \mathbf{G}_{t,t} \) incurs a complexity of \( O(n^2) \). Given that this multiplication is repeated \( k_a \) times, the overall complexity of this step is \( O(k_a \cdot n^2) \).
In the final phase, ML decoding is reapplied to the remaining 
\( N - k_a \) segments. The complexity for this phase is
\( O((N - k_a) \cdot f(n)) \) and the overall complexity \( C \) of the decoding algorithm can be summarized as:
\begin{equation}
C = O(k_a \cdot f(n)) + O(k_a \cdot n^2) + O((N - k_a) \cdot f(n)).    
\end{equation}
By combining the terms associated with \( f(n) \), we can express the complexity as $C = O(f(n) \cdot N) + O(k_a \cdot n^2)\approx O(N \cdot f(n))$.
In this section, we designed a decoding strategy leveraging the semi-systematic generator matrix structure, utilizing quantization and successive cancellation for efficient lattice point reconstruction. This approach ensures full diversity and linear complexity in decoding, making it suitable for reliable communication in fading environments while remaining scalable for large dimensions.


\section{Numerical Results}\label{result}
\begin{figure}[tb]
    \centering
    \includegraphics[width=9cm,trim=30 0 30 40, clip]{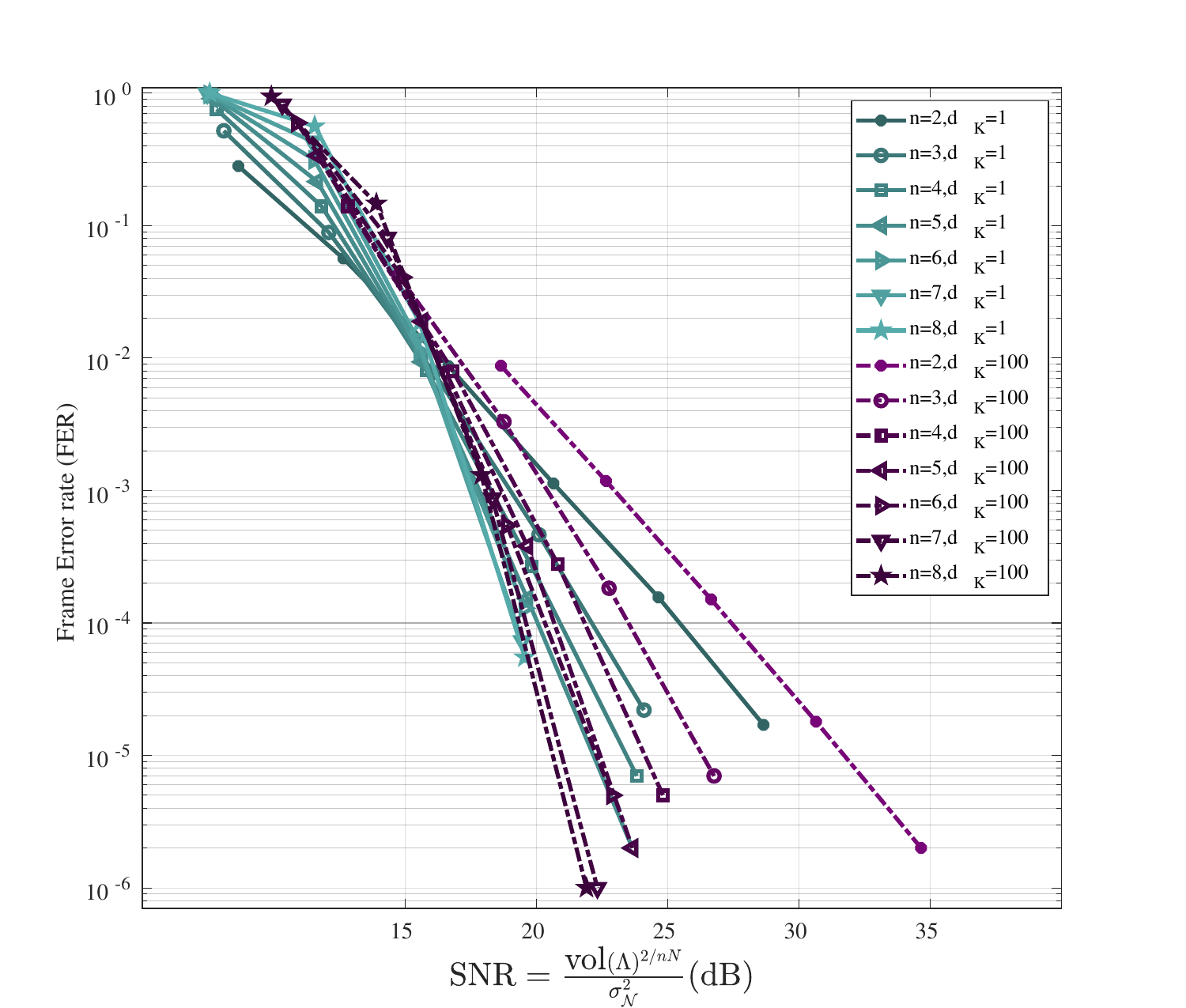}
    \caption{Poltyrev outage limit for full-diversity Construction-D lattices with $[N,k_2,k_1]=[100,50,40]$ and different diversity orders.}
    \label{comparison_polt}
\end{figure}

In this section, we present numerical results to assess the FER performance of the proposed decoding algorithm.  
The FER performance of all lattices is plotted against the SNR, defined as $\gamma = \frac{\text{vol}(\Lambda)^{\frac{2}{nN}}}{\sigma_{\mathcal{N}}^2}$. We compare our findings with the POL from \cite{POLlimit} and the performance of full-diversity Construction-A lattices presented in \cite{Hassan}.
This POL depends on the fading distribution and the determinant of the lattice, which in turn is influenced by 
$\Delta_K$ and the rate of its underlying codes. \figurename~\ref{comparison_polt} presents the Poltyrev outage limits of full-diversity  Construction-D lattices with varying parameters and diversity orders.

\begin{figure}[tb]
    \centering
    \includegraphics[width=9cm,trim=30 0 30 40, clip]{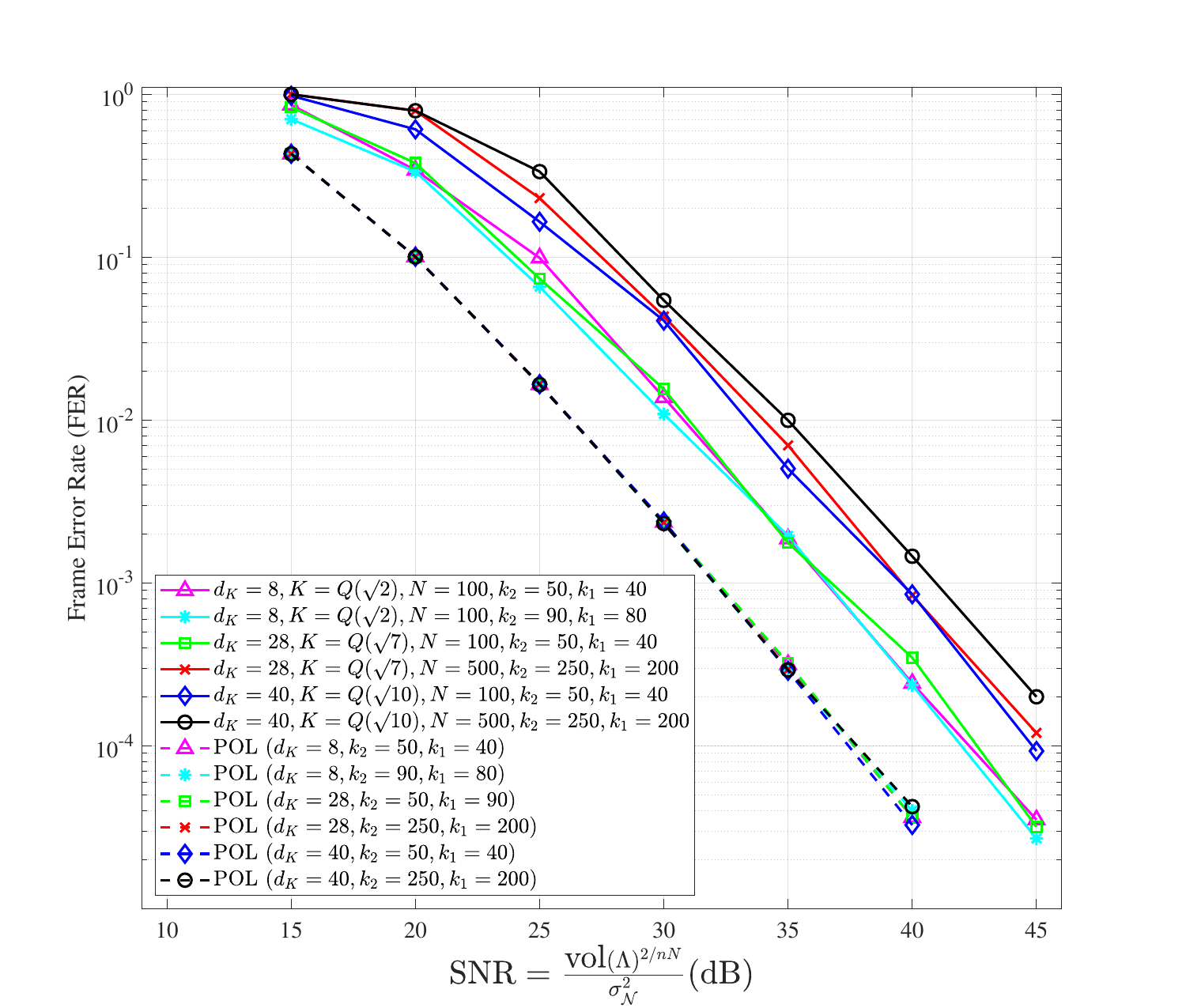}
    \caption{Decoding of diversity-2  Construction-D lattices with $m=2, 7, 10$.}
    \label{a_2}
\end{figure}

In Fig.~\ref{a_2}, the decoding results for double-diversity algebraic lattices are presented. The simulations use Construction-D lattices with $m = 2$, $m = 7$, and $m = 10$, decoded using the algorithm outlined in Section~\ref{decoding}. For \( m = 2 \) and \( m = 10 \), where \( m \equiv 2 \pmod{4} \), we obtain \( d_K = 4m = 8 \) and \( d_K = 40 \), respectively. In these scenarios, we consider \( K = \mathbb{Q}(\sqrt{m}) \) with \( \mathcal{O}_K = \mathbb{Z}[\sqrt{m}] \) and generator matrix $\mathbf{T}_1$. The prime ideals are  \( \mathfrak{P}_1 = 2\mathcal{O}_K + \sqrt{m}\mathcal{O}_K \) and \( \mathfrak{P}_2 = 2\mathcal{O}_K + 2\sqrt{m}\mathcal{O}_K \)  when \( a=2 \). The integral basis for \( \mathfrak{P}_1 \) is \(\{2, \sqrt{m}\}\), while for \( \mathfrak{P}_2 \), it is \(\{2, 2\sqrt{m}\}\). Consequently, the  utilized matrices \( \mathbf{T}_i \)'s in Algorithm~\ref{alg}, for \( i =1, 2, 3 \), are $\mathbf{T}_1 = \begin{bmatrix} 1 & 1 \\ \sqrt{m} & -\sqrt{m} \end{bmatrix}$, $\mathbf{T}_2 = \begin{bmatrix} 2 & 2 \\ \sqrt{m} & -\sqrt{m} \end{bmatrix}$
and $\mathbf{T}_3 = \begin{bmatrix} 2 & 2 \\ 2\sqrt{m} & -2\sqrt{m} \end{bmatrix}$.
For \( m = 7 \), where \( m \equiv 3 \pmod{4} \), we find \( d_K = 4m = 28 \). Here, \( K = \mathbb{Q}(\sqrt{7}) \), \( \mathcal{O}_K = \mathbb{Z}[\sqrt{7}] \), and the prime ideals are  \( \mathfrak{P}_1 = 2\mathcal{O}_K + (\sqrt{7} + 1)\mathcal{O}_K \) and \( \mathfrak{P}_2 = 2\mathcal{O}_K + 2\sqrt{7} \mathcal{O}_K \). In this case the integral basis for $\mathcal{O}_K$ is $\{1,\sqrt{7}\}$, for \( \mathfrak{P}_1 \) is \( \{2, 1 + \sqrt{7}\} \), and for \( \mathfrak{P}_2 \), it is \( \{2, 2\sqrt{7}\} \). Consequently, we have $\mathbf{T}_1 = \begin{bmatrix} 1 & 1 \\  \sqrt{7} & - \sqrt{7} \end{bmatrix}$, $\mathbf{T}_2 = \begin{bmatrix} 2 & 2 \\ 1 + \sqrt{7} & 1 - \sqrt{7} \end{bmatrix}$ and $\mathbf{T}_3 = \begin{bmatrix} 2 & 2 \\ 2\sqrt{7} & -2\sqrt{7} \end{bmatrix}$.

In \figurename~\ref{a_2}, at a FER of \(10^{-4}\), the double-diversity Construction-D lattice derived from \(Q(\sqrt{10})\) with parameters \([N, k_2, k_1] = [100, 50, 40]\) operates at a distance of $7$ dB from its corresponding POL. In contrast, the lattice presented in \cite{Hassan} performs  $8.45$ dB away from its POL. This reflects an improvement of $1.459$ dB over the results in \cite{Hassan}. The double-diversity Construction-D lattice constructed from \( Q(\sqrt{7}) \) with parameters \([N, k_2, k_1] = [100, 50, 40]\) achieves a distance of $4.8$ dB from its corresponding POL, surpassing the \( Q(\sqrt{10}) \)-based lattice by $2.2$ dB. Similarly, the lattice constructed over \( \mathbb{Q}(\sqrt{2}) \) with the same parameters demonstrates a performance gap of \( 4.5\) dB relative to its POL. In comparison, the Construction-A lattice over \( \mathbb{Q}(\sqrt{2}) \) with parameters \( [N, k] = [100, 50] \) shows a larger gap of \( 5.5 \) dB from its corresponding POL.
 This improved performance is due to the lower discriminants of \( Q(\sqrt{2}) \) and \( Q(\sqrt{7}) \) compared to \( Q(\sqrt{10}) \). 

\begin{figure}[tb]
    \centering
    \includegraphics[width=9cm,trim=30 0 30 40, clip]{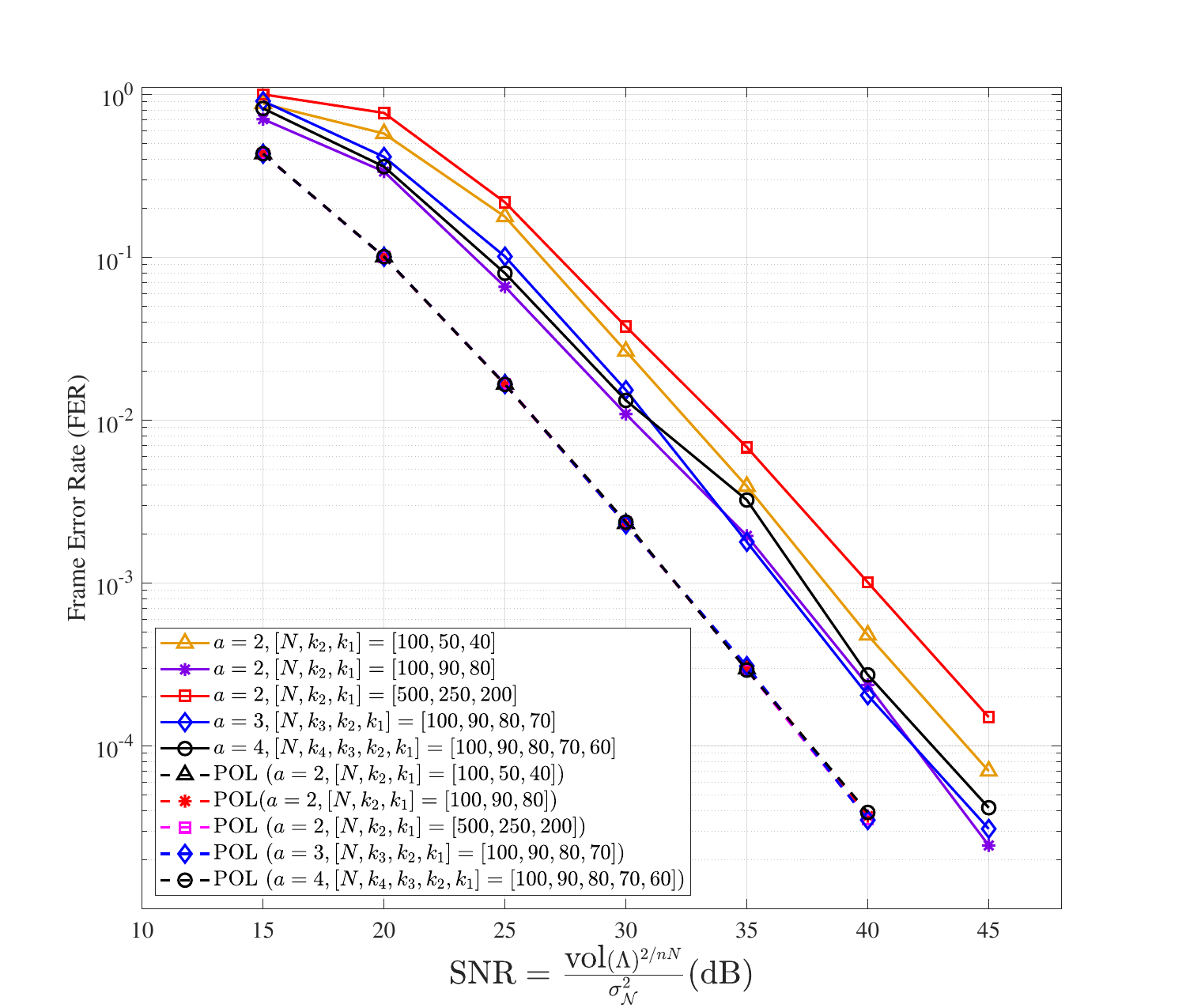}
    \caption{Decoding of diversity-2 Construction-D lattices over $\mathbb{Q}(\sqrt{2})$ for $a=2,3,4$.}
    \label{comparison_layers}
    \vspace{-0.5cm}
\end{figure}

 In \figurename~\ref{comparison_layers}, we compare the FER of double-diversity Construction-D lattices over \( \mathbb{Q}(\sqrt{2}) \) constructed with 2, 3, and 4 nested codes. It can be checked that augmenting the number of nested codes detrimentally impacts performance, whereas an increase in the code rate enhances it.
 At a FER of \(10^{-4}\), the lattice with parameters \([N, k_4, k_3, k_2, k_1] = [100, 90, 80, 70, 60]\) is \(4.9\) dB from its POL, while \([N, k_3, k_2, k_1] = [100, 90, 80, 70]\) and \([N, k_2, k_1] = [100, 90, 80]\) achieve \(4.3\) dB and \(4.1\) dB from their corresponding POL, respectively.

\begin{figure}[!htbp]
    \centering
    \includegraphics[width=9cm,trim=30 0 30 40, clip]{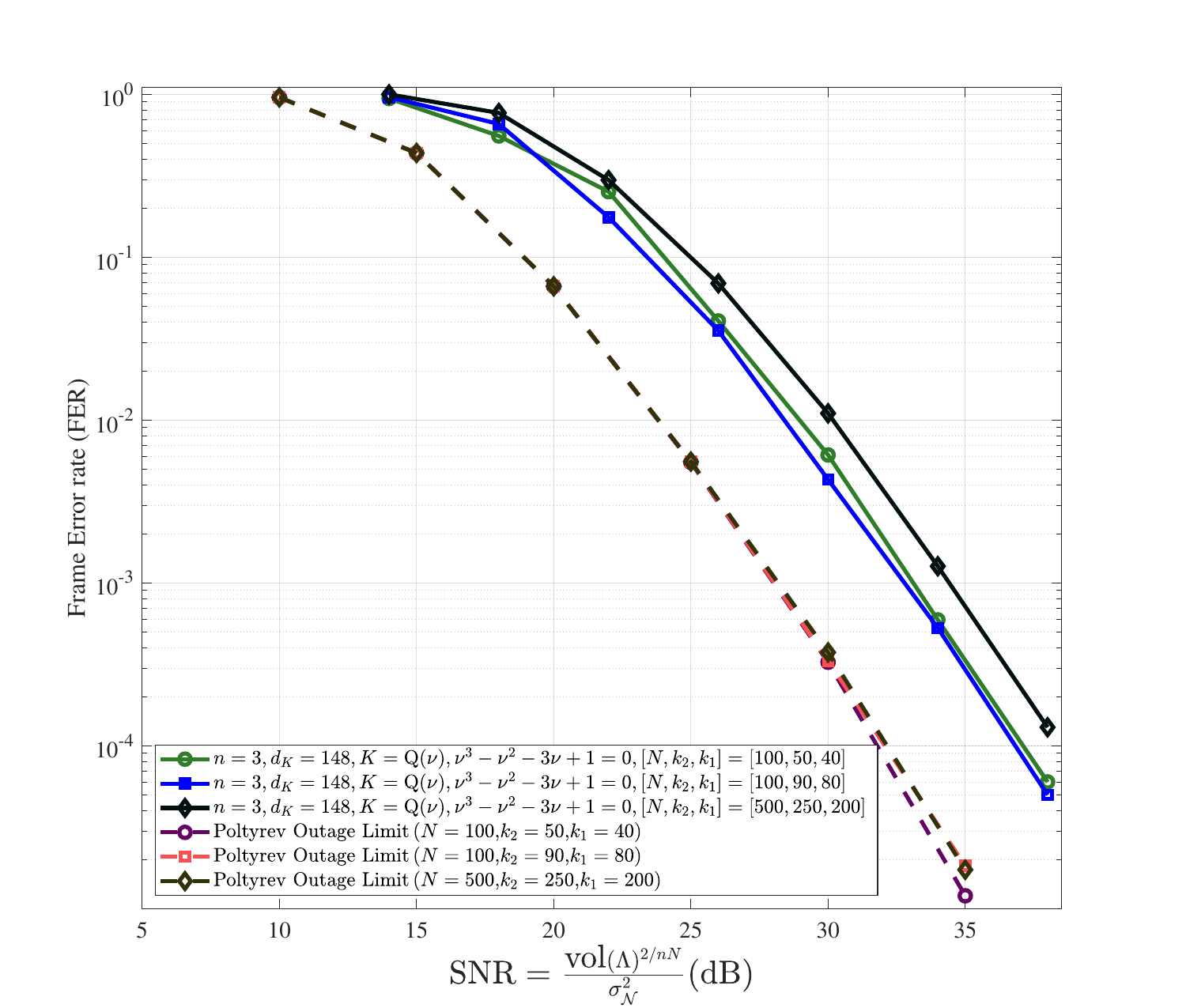}
    \caption{Decoding of diversity-3 Construction-D lattices over the number field $\mathbb{Q}(\nu)$ with discriminant $d_K=148$.}
    \label{triple}
\end{figure}

\setlength{\tabcolsep}{3pt} 
\renewcommand{\arraystretch}{0.9} 
\begin{table}[h]
    \centering
    \caption{Performance results of full-diversity Construction-D and Construction-A lattices. }
    \begin{tabular}{@{}cccccc@{}}
        \toprule
        \textbf{Lattice} & $(K,d_K)$ & $n$ & $a$ & $[N,k_1,\ldots,k_a]$ &\textbf{FER$= 10^{-4}$}  \\
        \midrule
       Construction-D & $(\mathbb{Q}(\sqrt{2}),8) $ &$2$  & $2$ & $[100,50,40]$ &   $4.50$ dB         \\
       Construction-D & $(\mathbb{Q}(\sqrt{7}),28) $ &$2$  & $2$ &  $[100,50,40]$ & $4.80$ dB  \\
       Construction-D & $(\mathbb{Q}(\sqrt{10}),40) $ &$2$  & $2$ & $[100,50 , 40]$&   $ 7.00$ dB \\
       Construction-D & $(\mathbb{Q}(\sqrt{7}),28) $ &$2$  & $2$ & $[500, 250, 200]$ & $8.10$ dB   \\
       Construction-D & $(\mathbb{Q}(\sqrt{10}),40) $ &$2$  & $2$ &$[500, 250, 200] $&  $8.40$ dB  \\
       Construction-D & $(\mathbb{Q}(\sqrt{2}),8) $ &$2$  & $2$ &$[500, 250, 200]$ & $7.90$ dB \\
       Construction-D & $(\mathbb{Q}(\sqrt{2}),8) $ &$2$  & $2$ &$[100, 90, 80]$& $4.10$ dB \\
       Construction-D & $(\mathbb{Q}(\sqrt{2}),8) $ &$2$  & $3$ &$[100, 90, 80, 70]$& $4.30$ dB  \\
       Construction-D & $(\mathbb{Q}(\sqrt{2}),8) $ &$2$  & $4$ &$[100, 90, 80, 70, 60]$&$4.90$ dB       \\
       Construction-D & $(\mathbb{Q}(\nu),148) $ &$3$  & $2$ &$[100,90 ,80] $& $4.80$ dB\\
       Construction-D & $(\mathbb{Q}(\nu),148) $ &$3$  & $2$ &$[100,50 ,40]$ & $5.10$ dB  \\
       Construction-D & $(\mathbb{Q}(\nu),148) $ &$3$  & $2$ &$[500,250 ,200]$ &  $ 6.40$ dB\\
       \hline\hline
       Construction-A & $(\mathbb{Q}(\sqrt{2}),8) $ &$2$  & $1$ & $[100,50]$ &   $5.50$ dB         \\
       Construction-A & $(\mathbb{Q}(\sqrt{7}),28) $ &$2$  & $1$ &  $[100,50]$ & $7.25$ dB  \\
       Construction-A & $(\mathbb{Q}(\sqrt{10}),40) $ &$2$  & $1$ & $[100,50]$&   $ 8.45$ dB \\
       Construction-A & $(\mathbb{Q}(\sqrt{10}),40) $ &$2$  & $1$ &$[500, 250] $&  $8.60$ dB  \\
       Construction-A & $(\mathbb{Q}(\sqrt{2}),8) $ &$2$  & $1$ &$[100, 90]$& $4.20$ dB \\
       Construction-A & $(\mathbb{Q}(\nu),148) $ &$3$  & $1$ &$[100 ,50]$ & $3.65$ dB  \\
       Construction-A & $(\mathbb{Q}(\nu),148) $ &$3$  & $1$ &$[500,250]$ &  $ 4.30$ dB\\
        \bottomrule
    \end{tabular}
    \label{tab:lattice_parameters}
\end{table}

In \figurename~\ref{triple}, we present the FER performance of triple-diversity Construction-D lattices constructed from binary nested codes with parameters \([N, k_2, k_1] = [100, 50, 40]\), \([100, 90, 80]\), and \([500, 250, 200]\). For comparison, the POL corresponding to a diversity order of three is also included. 
The lattice with parameters \([N, k_2, k_1] = [100, 50, 40]\) exhibits a gap of $5.1$~dB from its POL. This gap can be reduced by employing higher-rate underlying codes and selecting number fields with lower discriminants, both of which enhance the lattice's error-correcting capabilities. For the lattice with parameters \([100, 90, 80]\), corresponding to a lattice of dimension 300, the gap narrows slightly to $4.8$~dB. Lastly, the lattice with parameters \([500, 250, 200]\) corresponding to a lattice dimension of 1500 demonstrates a gap of $6.4$~dB from its POL, showing an increase in the performance gap compared to the lower-dimensional cases.
In \tablename~\ref{tab:lattice_parameters}, we present a comparative analysis of Construction-D lattices and their corresponding Construction-A lattices, including their respective parameters. The last column in this table  indicates the distance from POL at FER of $10^{-4}$.

These results highlight the interplay between lattice parameters, the rate of the underlying codes, and the properties of the number field on performance relative to theoretical limits. The decoding strategy also significantly influences the observed performance gap. Successive cancellation (SC) decoding, while computationally efficient, is prone to error propagation. Errors occurring during the early stages of SC decoding affect subsequent decoding stages, degrading the accuracy of the \(N - k_a\) points. Employing ML decoding for the first \(k_a\) points can reduce the initial error rate, but it does not fully eliminate the cascading effects of errors that propagate during the SC decoding of the remaining points.
\subsection{Future Research Directions}
The primary objective of this paper was to enhance the error performance of full-diversity Construction-A lattices by incorporating additional component codes and layers. This approach is analogous to the improved performance observed in lattices over AWGN channels, where increased layering yields better error performance. However, as shown in Table \ref{tab:lattice_parameters}, this improvement does not extend to lattices over BF channels. For these, full-diversity Construction-D lattices surpass full-diversity Construction-A lattices only in scenarios with two coding levels and diversity of 2. Unraveling the core reasons for this phenomenon remains an intriguing open problem.
Furthermore, developing a decoder that mitigates the error propagation inherent in SC decoding across higher layers, thereby enhancing error performance for all diversity orders, poses another fascinating challenge. Additionally, exploring the application of these novel Construction-D lattices in wiretap channels, and determining the necessary and sufficient conditions for them to be unimodular, is a prospective area for further research.
Finally, the construction and secrecy gain analysis of $d$-modular Construction-D lattices, derived from totally real and totally imaginary quadratic number fields for $d = 1, 3, 5, 6, 7, 11, 14, 15, 23$, is of significant interest for information security applications.

\section{Conslusion}\label{conclusion}
This work examined the application of algebraic lattices derived from Construction-D for communication over BF channels. We introduced a framework for defining Construction-D lattices, including their semi-systematic generator matrix, using nested linear codes and prime ideals from algebraic number fields. A decoding technique with linear complexity growth relative to the lattice dimension was also proposed, ensuring computational efficiency. Simulation results indicate that Construction-D lattices demonstrate a significantly lower Frame Error Rate (FER) compared to Construction-A lattices in scenarios involving diversity-2 and two-layer coding. These results establish Construction-D lattices as a powerful and efficient coding strategy for BF channels, providing significant improvements in communication system performance. Future research could focus on refining decoding methods, extending the applicability of these lattices to other channel conditions, and exploring hardware implementations to fully realize their practical potential.

\bibliographystyle{ieeetr} 
\bibliography{reference}

\begin{thebibliography}{10}

\bibitem{ISIT2025}
M.~Sadeghi, H.~Khodaiemehr, and C.~Feng, ``Design and decoding of full-diversity construction-d lattices on block-fading channels,'' in {\em 2025 IEEE International Symposium on Information Theory (ISIT)}, 2025.
\newblock Submitted for publication.

\bibitem{Poltyrev}
G.~Poltyrev, ``On coding without restrictions for the {AWGN} channel,'' {\em IEEE Trans. on Inf. Theory}, vol.~40, no.~2, pp.~409--417, 1994.

\bibitem{urbanke}
R.~Urbanke and B.~Rimoldi, ``Lattice codes can achieve capacity on the {AWGN} channel,'' {\em IEEE Trans. on Inf. Theory}, vol.~44, no.~1, pp.~273--278, 1998.

\bibitem{erez2004achieving}
U.~Erez and R.~Zamir, ``Achieving $\frac{1}{2} \log (1+ \mathrm{SNR})$ on the {AWGN} channel with lattice encoding and decoding,'' {\em IEEE Trans. on Inf. Theory}, vol.~50, no.~10, pp.~2293--2314, 2004.

\bibitem{conway1998nja}
J.~H. Conway and N.~J.~A. Sloane, {\em Sphere Packings, Lattices and Groups}.
\newblock New York: Springer-Verlag, 3rd~ed., 1998.

\bibitem{barnes1983new}
E.~Barnes and N.~Sloane, ``New lattice packings of spheres,'' {\em Canadian Journal of Mathematics}, vol.~35, no.~1, pp.~117--130, 1983.

\bibitem{forney2000sphere}
G.~D. Forney, M.~D. Trott, and S.-Y. Chung, ``Sphere-bound-achieving coset codes and multilevel coset codes,'' {\em IEEE Trans. on Inf. Theory}, vol.~46, no.~3, pp.~820--850, 2000.

\bibitem{1057135}
J.~Conway and N.~Sloane, ``Soft decoding techniques for codes and lattices, including the {G}olay code and the {L}eech lattice,'' {\em IEEE Trans. on Inf. Theory}, vol.~32, no.~1, pp.~41--50, 1986.

\bibitem{LDLC}
N.~Sommer, M.~Feder, and O.~Shalvi, ``Low-density lattice codes,'' {\em IEEE Trans. on Inf. Theory}, vol.~54, pp.~1561--1585, Apr. 2008.

\bibitem{LDA}
N.~di~Pietro, J.~J. Boutros, G.~Z\'{e}mor, and L.~Brunel, ``Integer low-density lattices based on {C}onstruction \textsc{A},'' in {\em IEEE Inf. Theory Workshop (ITW), 2012}, pp.~422--426, Sept. 2012.

\bibitem{LDA2}
N.~di~Pietro, J.~J. Boutros, G.~Z\'{e}mor, and L.~Brunei, ``New results on low-density integer lattices,'' in {\em Inf. Theory and Applications Workshop (ITA), 2013}, pp.~1--6, Feb. 2013.

\bibitem{LDA3}
N.~di~Pietro, G.~Z\'{e}mor, and J.~J. Boutros, ``New results on {C}onstruction \textsc{A} lattices based on very sparse parity-check matrices,'' in {\em IEEE Int. Symp. on Inf. Theory (ISIT), 2013}, pp.~1675--1679, Jul. 2013.

\bibitem{Leech}
N.~{di Pietro} and J.~J. {Boutros}, ``Leech constellations of {C}onstruction-{A} lattices,'' {\em IEEE Trans. on Commun.}, vol.~65, pp.~4622--4631, Nov. 2017.

\bibitem{polar}
Y.~Yan and C.~Ling, ``A construction of lattices from polar codes,'' in {\em IEEE Inf. Theory Workshop (ITW), 2012}, pp.~124--128, Sept. 2012.

\bibitem{QC_LDPC_lattice}
H.~Khodaiemehr, M.-R. Sadeghi, and A.~Sakzad, ``Practical encoder and decoder for power constrained {QC LDPC}-lattice codes,'' {\em IEEE Trans. on Commun.}, vol.~65, pp.~486--500, Feb. 2017.

\bibitem{matsumine2018construction}
T.~Matsumine, B.~M. Kurkoski, and H.~Ochiai, ``Construction {D} lattice decoding and its application to {BCH} code lattices,'' in {\em 2018 IEEE Global Commun. Conference (GLOBECOM)}, pp.~1--6, IEEE, 2018.

\bibitem{Liu}
L.~Liu, Y.~Yan, C.~Ling, and X.~Wu, ``Construction of capacity-achieving lattice codes: Polar lattices,'' {\em IEEE Trans. on Commun.}, vol.~67, no.~2, pp.~915--928, 2019.

\bibitem{blockfading}
L.~H. Ozarow, S.~Shamai, and A.~D. Wyner, ``Information theoretic considerations for cellular mobile radio,'' {\em IEEE Trans. on Vehicular Technology}, vol.~43, pp.~359--378, May 1994.

\bibitem{rootLDPC}
J.~J. Boutros, A.~Guill\'{e}n~i F\`{a}bregas, E.~Biglieri, and G.~Z\'{e}mor, ``Low-density parity-check codes for nonergodic block-fading channels,'' {\em IEEE Trans. on Inf. Theory}, vol.~56, pp.~4286--4300, Sept. 2010.

\bibitem{8187356}
F.~Oggier and E.~Viterbo, {\em Algebraic Number Theory and Code Design for {R}ayleigh Fading Channels}.
\newblock (Foundation and Trends in Commun. and Inf. Theory), 2004.

\bibitem{ebeling}
W.~Ebeling, {\em Lattices and Codes: A Course Partially Based on Lectures by {Friedrich Hirzebruch}}.
\newblock Advanced Lectures in Mathematics, New York: Springer Fachmedien Wiesbaden, 2012.

\bibitem{ConstA}
W.~Kositwattanarerk, S.~S. Ong, and F.~Oggier, ``Construction \textsc{A} of lattices over number fields and block fading (wiretap) coding,'' {\em IEEE Trans. on Inf. Theory}, vol.~61, pp.~2273--2282, May 2015.

\bibitem{Hassan}
H.~Khodaiemehr, D.~Panario, and M.-R. Sadeghi, ``Design and practical decoding of full-diversity {C}onstruction {A} lattices for block-fading channels,'' {\em IEEE Trans, on Inf. Theory}, vol.~67, no.~1, pp.~138--163, 2021.

\bibitem{strey2017lattices}
E.~Strey and S.~I. Costa, ``Lattices from codes over $ \mathbb{Z}_q$: generalization of {C}onstructions {D}, {D}$'$ and $\bar{D}$,'' {\em Designs, Codes and Cryptography}, vol.~85, pp.~77--95, 2017.

\bibitem{do2023lattice}
F.~do~Carmo~Silva, A.~P. de~Souza, E.~Strey, and S.~I. Costa, ``On lattice {C}onstructions {D} and {D}$'$ from $q$-ary linear codes,'' {\em Commun. in Mathematics}, vol.~31, 2023.

\bibitem{485720}
J.~Boutros, E.~Viterbo, C.~Rastello, and J.-C. Belfiore, ``Good lattice constellations for both {R}ayleigh fading and {G}aussian channels,'' {\em IEEE Trans. on Inf. Theory}, vol.~42, no.~2, pp.~502--518, 1996.

\bibitem{Alaca}
S.~Alaca and K.~S. Williams, {\em Introductory Algebraic Number Theory}.
\newblock Cambridge University Press, 2003.

\bibitem{serglang}
S.~Lang, {\em Algebraic Number Theory}.
\newblock Springer-Verlag, 1994.

\bibitem{Stewart}
I.~N. Stewart and D.~O. Tall, {\em Algebraic Number Theory}.
\newblock Chapman and Hall, 1979.

\bibitem{sahai2003algebra}
V.~Sahai and V.~Bist, {\em Algebra}.
\newblock Alpha Science International, 2003.

\bibitem{7282492}
R.~Vehkalahti and L.~Luzzi, ``Number field lattices achieve {G}aussian and {R}ayleigh channel capacity within a constant gap,'' in {\em 2015 IEEE Int. Symp. on Inf. Theory (ISIT)}, pp.~436--440, 2015.

\bibitem{POLlimit}
M.~Punekar, J.~J. Boutros, and E.~Biglieri, ``A {Poltyrev} outage limit for lattices,'' in {\em 2015 IEEE Int. Symp. on Inf. Theory (ISIT)}, pp.~456--460, 2015.

\end{thebibliography}
\end{document}